\documentclass[acmsmall]{acmart}

\usepackage{amsmath, amsfonts}
\usepackage{makecell}
\usepackage{subcaption}
\usepackage{tikz}
\usepackage{listings}
\usepackage{svg}
\usepackage{color}
\usepackage{float}
\usepackage{overpic} 
\usepackage{multirow}
\usepackage{caption}
\usepackage{xcolor}
\usepackage{algpseudocode}
\usepackage{dashrule}
\usepackage{booktabs}
\usepackage{graphicx}
\usepackage{enumitem}
\usepackage{colortbl}
\usepackage{tcolorbox}
\usepackage{xspace}
\usepackage{circledtext}
\usepackage{fontawesome5}
\usepackage{tabularx} 
\usepackage{booktabs} 
\usepackage{multirow}

\newcommand{\xzp}[1]{{\color{black} {#1}}}
\newcommand{\xt}[1]{{\color{black} {#1}}}
\newcommand{\tool}{\textsc{\textbf{SmellCC}}\xspace}

\newcommand{\find}[1]{
\begin{tcolorbox}[leftrule=0.5mm,rightrule=0.5mm, toprule=0.5mm,bottomrule=0.5mm,left=2pt,right=2pt,top=2pt,bottom=2pt]
\em #1
\end{tcolorbox}
}

\AtBeginDocument{%
  }

\setcopyright{acmlicensed}
\copyrightyear{2018}
\acmYear{2018}
\acmDOI{XXXXXXX.XXXXXXX}

\acmJournal{JACM}
\acmVolume{37}
\acmNumber{4}
\acmArticle{111}
\acmMonth{8}

\begin{document}

\title{Clean Code, Better Models: Enhancing LLM Performance with Smell-Cleaned Dataset}

\author{Zhipeng Xue}
\authornote{Both authors contributed equally to this research.}
\email{zhipengxue@zju.edu.cn}
\author{Xiaoting Zhang}
\authornotemark[1]
\email{xiaotingzhang@zju.edu.cn}
\affiliation{%
  \institution{Zhejiang University}
  \city{HangZhou}
  \country{China}
}

\author{Zhipeng Gao}
\authornote{Corresponding author.}
\affiliation{%
  \institution{Shanghai Institute for Advanced Study of Zhejiang University}
  \city{Shanghai}
  \country{China}
}
\email{zhipeng.gao@zju.edu.cn}

\author{Xing Hu}
\authornotemark[2]
\affiliation{%
  \institution{Zhejiang University}
  \city{HangZhou}
  \country{China}
}
\email{xinghu@zju.edu.cn}

\author{Shan Gao}
\affiliation{%
  \institution{Independent Researcher}
  \city{HangZhou}
  \country{China}
}
\email{gaoshan17@huawei.com}

\author{Xin Xia}
\affiliation{%
  \institution{Zhejiang University}
  \city{HangZhou}
  \country{China}
}
\email{xin.xia@acm.org}

\author{Shanping Li}
\affiliation{%
  \institution{Zhejiang University}
  \city{HangZhou}
  \country{China}
}
\email{shan@zju.edu.cn}


\begin{abstract}
The Large Language Models (LLMs) have demonstrated great potential in code-related tasks. 
However, most research focuses on improving the output quality of LLMs (e.g., correctness), and less attention has been paid to the LLM input (e.g., the training code quality). 
Given that code smells are widely existed in practice and can negatively impact software maintainability and readability, this study takes the first systematic research to assess and improve dataset quality in terms of code smells. 
In this work, we first conduct a preliminary study to explore the presence of code smells in a popular benchmark dataset (i.e., \xt{\texttt{CodeSearchNet}-Python}) and evaluate the output of several popular LLMs (i.e., DeepSeek-Coder, CodeLlama, and MagiCoder), revealing that code smell issues extensively exist in LLM's input (e.g., benchmark dataset) and output (e.g., generated code). 
We also perform a user study to investigate developers' perspectives on LLM-generated code with and without smells, which indicated developers' strong preference for smell-free code and their willingness to leverage LLMs for intricate code smell removal.
We then conduct our systematic research by taking three main steps: Firstly, we propose an LLM-based code smell cleaning tool, named \tool (\textbf{\underline{Smell}} \textbf{\underline{C}}ode \textbf{\underline{C}}leaner), which automatically refactors and removes code smells. 
\xt{
To evaluate the correctness of the code refactoring, we construct a test set of 50 repositories sourced from the \texttt{CodeSearchNet}-Python benchmark for functional testing.
Then we apply our curated smell-cleaned dataset to fine-tune two LLMs (i.e., DeepSeek-V2 and Qwen-Coder) to explore their potential for generating high-quality code. 
Thirdly, we investigate the impact of code smells on two downstream tasks: code completion and code search. 
Lastly, we derive several actionable implications for software engineering researchers and industry practitioners from our findings.
The experimental results show that our \tool eliminates 91.6\% of code smells across the entire \texttt{CodeSearchNet}-Python corpus, curating a smell-cleaned benchmark dataset. On a curated 50-repository subset, \tool achieves 96.8\% smell removal while maintaining 91.3\% correctness through test verification.
The LLMs fine-tuned on this smell-cleaned dataset reduce code smells in generated code by 79.6\% and 83.1\% for DeepSeek-V2 and Qwen-Coder respectively. 
Moreover, applying the smell-cleaned dataset to code completion and code search tasks yields significant improvements across all models (DeepSeek-V1/V2, Qwen-Coder), with Qwen-Coder achieving peak gains of 12.2\% in completion and 4.3\% in search performance.
}
\end{abstract}



\received{20 February 2007}
\received[revised]{12 March 2009}
\received[accepted]{5 June 2009}

\maketitle

\section{INTRODUCTION}
\textit{Code smells} (or shortly ``smells'') refers to the poor design or implementation that may adversely affect the maintenance of the software. 
Code smells are often considered as bad practices performed by developers in development, such as dead code and long method/parameter lists, they are not bugs or errors but are observable violations of code design/development fundamentals that could eventually lead to poor code quality and technical debt. 
In other words, the code may compile and execute as expected, but code smells can hurt the software maintainability in the long term and cause performance or security issues in the future~\cite{10006873,santos2018systematic, Pereira_dos_Reis_2021,lacerda2020code, wang2024just}. 

\begin{figure}
    \centering
    \includegraphics[width=0.95\textwidth]{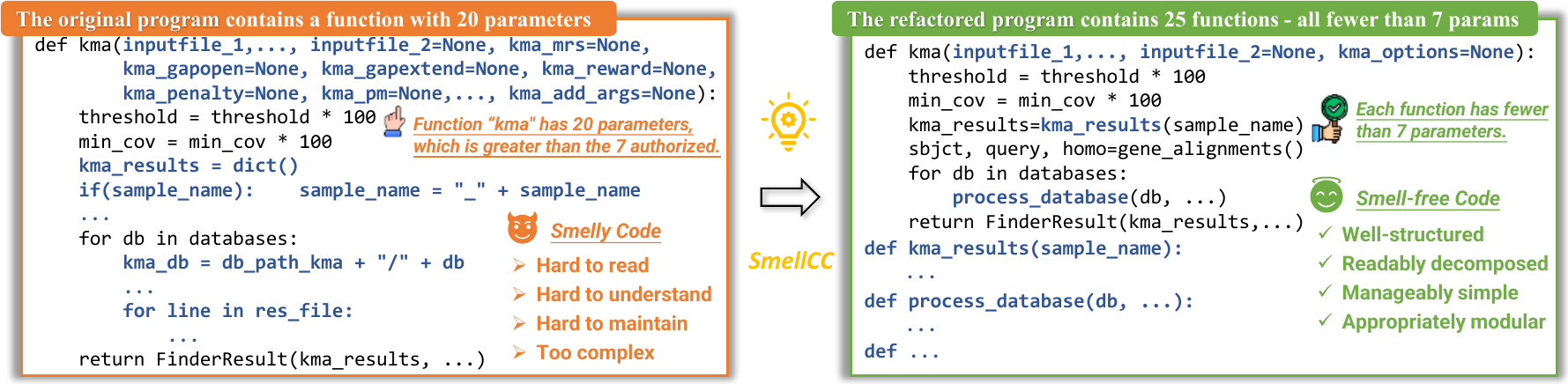} 
    \caption{An Example of Long Parameter List Smell}
    \label{fig:long method}
\end{figure}

Fig.~\ref{fig:long method} demonstrates an example of the \textit{Long Parameter List} smell. 
This code smell occurs when a method grows too long. 
When a method gets too long, suggesting this function is responsible for too many tasks. 
The \textit{Long Parameter List} smell makes the code difficult to read, test and/or reuse. 
\xt{This smell can be cleaned by encapsulating the parameters within a dedicated dataclass structure or breaking the long method into small sub-methods, where each handles a single task.} 
Cleaning this code smell can make the code more maintainable and easy to read. 
These code smells should be prevented and cleaned from the code base as early as possible. 
However, in practice, developers usually lack the time and resources to avoid code smells from happening. 
As a result, these code smells continue to stay in the system and increase the system complexity over time.

Recently, large language models (LLMs) have demonstrated great potential for assisting developers in their daily development~\cite{svyatkovskiy2020intellicode, ryan2024code, jin2023inferfix, liu2024non, qiu2021deep, xue2024selfpico, gao2024automating, yan2023closer}. 
The effectiveness of LLMs heavily relies on the quality of the data they are trained on~\cite{shi2022we, croft2023data, gao2024learning, su2023still, cote2024data, wang2024makes, gao2020technical, yang2024federated}. 
So far, most research focuses on improving LLM's output quality (e.g., how to generate correct code~\cite{liu2024your, li2023structured, Chen2024JumpCoderGB, dai2024mpcoder}), but less attention has been paid to the quality of the input data, such as the large-scale code corpus. 
Given code smells are unavoidable in practice and Code LLMs are trained with code samples from real-world software projects (e.g., GitHub), several important and interesting research questions are emerging that deserve further investigation: 
\circledtextset{resize=real}
\circledtext*[height=1.8ex]{1} \textit{Do code smells in the training data affect the quality of LLMs' output?} 
If LLMs are trained with ``\textit{smelly}'' code as input, do they also tend to produce code with similar issues? 
\circledtext*[height=1.8ex]{2} \textit{Can LLMs be utilized to clean different types of code smells while maintaining the same code functionality}?
Inspired by the great success of LLMs in code generation and code comprehension, is it possible to leverage LLMs to automatically refactor and clean code smells, thus curating a smell-cleaned dataset for further training?  
\circledtext*[height=1.8ex]{3} If so, \textit{do the smell-cleaned dataset can be used, in turn, to fine-tune LLMs to generate high-quality outputs?} 
If the smell-cleaned dataset is curated, can it potentially lead LLM to generate higher-quality and more maintainable code with fewer code smells? 
\circledtext*[height=1.8ex]{4} Finally, \textit{can the smell-cleaned dataset enhance the model's performance in other downstream tasks?} When the quality of the dataset is improved, can it benefit other downstream tasks?  

To investigate the aforementioned concerns of data quality for LLMs, in this work, we conduct a systematic research to improve and assess the dataset quality in terms of code smells. 
Specifically, we choose \texttt{CodeSearchNet} benchmark~\cite{husain2019codesearchnet} as our objective dataset. 
\texttt{CodeSearchNet} is a benchmark dataset collected from real-world GitHub projects, which consists of two million $\langle$comment, method$\rangle$ pairs of multiple programming languages (e.g., Python, Javascript, Ruby, Java, etc). 
Many LLMs, including CodeT5~\cite{wang2021codet5}, StarCoder~\cite{li2023starcoder}, and SantaCoder~\cite{allal2023santacoder}, are trained and/or evaluated on it. 
Despite \texttt{CodeSearchNet} being widely used for training LLMs, little is known about the quality of the benchmark dataset, especially from perspectives of code smells. 
Our preliminary study shows that over 200K code smells are detected (by using SonarQube~\cite{SonarQube}) in \texttt{CodeSearchNet}-Python dataset, and the same code smells are also detected within generated code of popular code LLMs (i.e., DeepSeek-Coder~\cite{DBLP:journals/corr/abs-2401-14196}, CodeLlama~\cite{roziere2023code} and MagiCoder~\cite{wei2023magicoder}), which confirms our assumption that \textbf{code smells extensively existed in benchmark datasets and they can indeed affect the quality of the LLM generated code}. 
Furthermore, our user study investigating developers' perceptions of code smells in LLM-generated code reveals that most developers are inclined to accept smell-free code and are generally receptive to using LLMs for eliminating code smells.
Following that, we conduct our systematic research by taking three main steps: Firstly, we build an LLM-based code smell cleaning tool, named \tool (\textbf{\underline{Smell}} \textbf{\underline{C}}ode \textbf{\underline{C}}leaner), for automatically refactoring and cleaning code smells in a given method. 
\xt{
The manual evaluation results show that 91.8\% of code smells can be successfully eliminated and removed, curating a smell-cleaned dataset for exploration. 
To verify the functional consistency of the code before and after refactoring, we construct a test set of 50 repositories sourced from the \texttt{CodeSearchNet}-Python corpus for testing, which suggests that \tool can effectively remove most of the code smells (i.e., 96.8\%) while maintaining the same functionality (i.e., 91.3\%).
Secondly, we fine-tuned the LLMs (i.e., DeepSeek-V2 and Qwen-Coder) with our curated smell-cleaned dataset, the experimental results show that compared with the original LLMs, code smells are significantly reduced in generated code after fine-tuning, improving the quality and maintainability of the LLMs' outputs. 
Thirdly, we investigated the impacts of code smells on two downstream tasks, i.e., code completion and code search. 
The experimental results show that, cleaning code smells have a positive influence on different downstream tasks. 
Training and/or evaluating with smell-cleaned data improve Qwen-Coder's performance by 12.2\% and 4.3\% on code completion and code search respectively. 
Lastly, we derive several actionable implications for software engineering researchers and industry practitioners from our findings.
}

In summary, this paper makes the following contributions: 
\begin{itemize}
    \item To the best of our knowledge, this is the first systematic empirical study that investigates the code smells in an existing benchmark dataset, their impact on LLMs' generated codes, and developers' perspectives on these codes with/without code smells, providing practical insights for future code smell research.
    \item We develop an LLM-based code smell cleaning tool, named \tool, designed to uncover LLMs' capability to remove code smells in a given method, which can help curate high-quality smell-cleaned dataset. 
    \item We conduct a comprehensive analysis on the performance of LLMs trained on the origin and smell-clean benchmark dataset, our experimental results show that cleaning code smells yields significant improvement in both LLM's generated code and downstream tasks. 
    \item We release \tool and the curate smell-cleaned benchmark dataset~\cite{replication_package}, in order to facilitate other researchers to replicate our study and verify their own ideas. 
\end{itemize}
\section{PRELIMINARIES}
In this section, we first perform a preliminary study to explore whether code smells are present in both the input (e.g., the training dataset) and the output (e.g., generated code) of LLMs. Then we carry out a user study to investigate developers' attitudes toward LLM-generated code, with and without code smells.

\subsection{Code Smells in LLMs' Training Dataset} 
LLMs require large-scale and high-quality training datasets, while low-quality datasets can potentially lead LLMs to generate code with quality issues. 
However, the quality of the widely used benchmark dataset is still unknown, especially in terms of code smells. 
Therefore, we first conduct a preliminary study to investigate whether code smells widely exist in-the-wild in the benchmark dataset. 
To do so, we choose \texttt{CodeSearchNet} dataset~\cite{husain2019codesearchnet} for this study. 
\texttt{CodeSearchNet} contains 2 million $\langle$\textit{comment}, \textit{code}$\rangle$ pairs collected from GitHub open-source projects across six programming languages (i.e., Python, Javascript, Ruby, Go, Java, and PHP). 
The code data is at the granularity of method-level, and the comment is a natural language description for the code. 
\texttt{CodeSearchNet} has been widely used as a benchmark dataset for training LLMs, such as CodeT5~\cite{wang2021codet5} and StarCoder~\cite{li2023starcoder}. 
In this study, we choose \texttt{CodeSearchNet}-Python dataset for subsequent experiments, comprising of 450K Python functions and their corresponding descriptions. 
After that, we use a popular static analysis tool, SonarQube~\cite{SonarQube}, to detect code smells in Python code. 
SonarQube is a widely used code smell detecting tool to improve code quality~\cite{marcilio2019static, lenarduzzi2020some}. 


\begin{table} 
  \centering
  \caption{Top-10 Code Smells in \texttt{CodeSearchNet}-Python Dataset}
  \label{tab:pre_code_smell}
  \scalebox{0.9}{
  \begin{tabular}{lp{7cm}c}
    \toprule
    Type & Description & Amount\\
    \midrule
    \multirow{2}{*}{\textit{Commented Code}} & Outdated or unused code is left commented out in the codebase. & \multirow{2}{*}{10,509} \\
    \midrule
    \multirow{2}{*}{\textit{Naming Convention}} & Local variables, parameters, fields, and method names should follow the naming standards. &  \multirow{2}{*}{133,323} \\
    \midrule
    \multirow{2}{*}{\textit{Empty Nested Code Blocks}} & Nested code blocks that do not contain other statements except pass statements. & \multirow{2}{*}{2,446}\\
    \midrule
    \multirow{2}{*}{\textit{Collapsible if Statements}} & Multiple, consecutive ``if'' statements with related or overlapping conditions. & \multirow{2}{*}{8,631} \\
    \midrule
    \multirow{2}{*}{\textit{Long Parameter List}} & A method or function is excessively lengthy, complex, or has too many parameters. & \multirow{2}{*}{12,967} \\
    \midrule
    \multirow{3}{*}{\textit{High Cognitive Complexity}} & Code that is difficult to understand due to its intricate logic, deep nesting, or numerous conditional branches.  & \multirow{3}{*}{33,926}\\
    \midrule
    \multirow{2}{*}{\textit{Dead Code}} & Code exists in the codebase but is never reached or used. & \multirow{2}{*}{458} \\
    \midrule
    \multirow{2}{*}{\textit{Self-assigned Variables}} & Variables that are identical on both sides of the assignment operator. & \multirow{2}{*}{458}\\
    \midrule
    \multirow{2}{*}{\textit{Identical Expressions}} & The same expressions are used on both sides of a binary operator. & \multirow{2}{*}{267}\\
    \midrule
    \multirow{2}{*}{\textit{Return and Yield}} & ``return'' and ``yield'' should not be used in the same function.  & \multirow{2}{*}{195}\\
    \bottomrule
  \end{tabular}
  }
\end{table}

The preliminary evaluation results are shown in Table~\ref{tab:pre_code_smell}, we demonstrate Top-10 code smells and their type description and occurrences in the table. 
From the table, we can see that: (1) \textbf{SonarQube successfully identified over 200K code smells across Top-10 code smell categories, which further verifies our assumption that code smells extensively exist in benchmark datasets.} 
We attribute the pervasive code smells in software projects to the following reasons: (a) In a fast-paced software development process, developers may resort to quick and dirty solutions that introduce code smells, particularly facing tight deadlines or complex requirements; (b) Less experienced developers or newcomers are not aware of the best practice, they may write code violating coding conventions; (c) Even experienced developers often choose to ignore code smells as they do not affect code functionality immediately;
(2) \textbf{These widely spread code smells pose threats to the quality of the dataset and risks for subsequent training processes}. 
For example, the \textit{Naming Convention} smell was detected over 133K times in the dataset, indicating inconsistent and unclear identifiers. 
\textit{Long Parameter List} and \textit{High Cognitive Complexity} smells make the code hard to maintain and understand. 
\textit{Commented Code} and \textit{Dead Code} smell present outdated and unused code within the dataset, which may potentially mislead LLMs with obsolete information. 
\textit{Collapsible if Statements} smell can hinder code performance and \textit{identical Expression} smell may introduce logical errors and even cause unexpected bugs in future.

\subsection{Code Smells in LLMs' Generated Code} 
LLMs require large-scale and high-quality datasets for training. 
These training datasets are usually constructed from real-world open-source software projects, e.g., GitHub software repositories~\cite{chang2024survey}. 
As discussed in Section 2.1, code smells extensively exist in open-source projects and pose threats to training dataset quality. 
\textit{Do code smells in the training dataset affect the quality of LLMs' output?} 
We thus further explore whether code smells also exist in LLMs' generated code. 
Specifically, we adopt three different open-source LLMs, i.e., DeepSeek-Coder~\cite{DBLP:journals/corr/abs-2401-14196}, CodeLlama~\cite{roziere2023code} and MagiCoder~\cite{wei2023magicoder}, for this preliminary study. 
We randomly sampled 1,000 natural language descriptions from \texttt{CodeSearchNet}-Python as our evaluation dataset. 
After that, we feed three LLMs with the same prompt (i.e., ``\textit{Please generate a Python function according to the following function description}'') to obtain its generated code. 
Each LLM generated 1k functions for evaluation. 
Lastly, SonarQube was applied to the generated 3k functions to identify potential code smells. 

The experimental results are shown in Table~\ref{tab:pre_gene_code}. 
From the table, we can see that: \textit{LLMs' generated code contains large number of code smells.}
This is reasonable because LLMs learned these patterns from the training datasets, where code smells widely existed. 
Nowadays, developers usually resort to LLMs for generating code during their daily development. 
When code smells occur in LLMs' generated code, developers have to carefully review the code and apply manual corrections or refactoring to remove the code smells. 
Therefore, the key question we ask in this work is: \textit{can we design models to automatically clean code smells in the training dataset, and ultimately enable LLMs to generate higher quality code?} 



\begin{table} 
\setlength{\abovecaptionskip}{5pt}
  \centering
  \caption{Code Smells in LLMs' Generated Code}
  \label{tab:pre_gene_code}
  \begin{tabular}{lccc}
    \toprule
    Type & CodeLlama & DeepSeek-Coder &MagiCoder \\
    \midrule
    \textit{Naming Convention} &93 & 116& 115\\
    \textit{High Cognitive Complexity} &9 &12 & 10\\
    \textit{Long Parameter List} & 12 & 9 & 11\\
    \textit{Collapsible if Statements} & 11 & 11 & 14\\
    \textit{Empty Nested Code Blocks} & 0 &3 & 3\\
    \textit{Self-assigned Variables} & 15 & 1 &0\\
    \textit{Identical Expressions} & 1 & 0 &0 \\
    \bottomrule
  \end{tabular}
\end{table}



\subsection{Developers' Perspectives on LLM-generated Code Smells}
As evidenced in Section 2.2, LLM-generated code exhibits a significant presence of code smells. 
So we further conducted a small-scale user study to gather developers’ attitudes toward LLMs’ generated code, both with and without code smells.

\xt{
Specifically, we conducted a questionnaire-based study with 10 experienced developers (each with over five years of programming experience). This study focused on seven types of code smells identified in Section 2.2. 
For each smell type, participants were sequentially presented with: (1) a description and an illustrative example of the smell, (2) an LLM-generated snippet containing the smell, and (3) a smell-free counterpart. 
Then, developers were asked to evaluate both smell and smell-free snippets per category without time constraints. 
For each snippet, developers should answer the following question: \textit{How would you interact with this code? (Accept/Reject/Refactor)}. If ``\textit{Refactor}'' was chosen, they would further be asked to decide: \textit{How would you refactor it? (Manually/Using static refactor tools/Using LLM/Other ways)}.
Such design allowed us to measure developers’ sensitivity to code smells in LLM outputs and their preferred remediation strategies.
}

The results revealed that developers directly accepted 83.3\% of code without code smells. 
Conversely, 76.7\% of code with code smells were marked for refactoring. 
Regarding their approach to refactor, developers preferred manual corrections for simpler code smells, such as \textit{Commented Code} (75\% opting for manual fixes). However, for more complex issues (e.g., reducing \textit{High Cognitive Complexity}), 71\% of developers preferred to leverage LLM for assistance. 
Developers’ feedback further motivates us to explore how to reduce code smells in LLMs’ generated code and how to use LLMs to assist code smell cleaning.

\find{\textbf{Preliminary Study Findings:} 
Code smells extensively exist in LLMs' training datasets, and at the same time, these code smells are also appear in LLMs' generated code, posing great threats to reusing the generated code in practice. What's more, most developers tend to accept codes without code smells and prefer to leverage LLMs for intricate code smell removal, which motivates our research on using LLMs to achieve code smell elimination and smell-free code generation.
}

\section{STUDY DESIGN}
\begin{figure}
    \centering
    \includegraphics[width=0.95\textwidth]{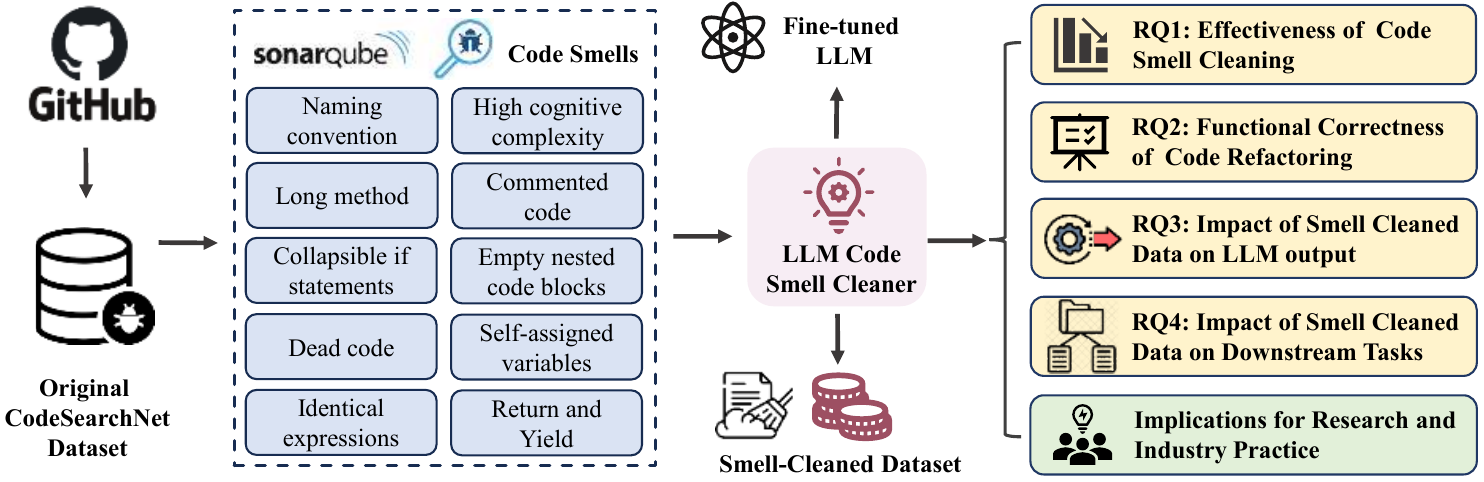}
    \caption{Overview of Our Study}
    \label{fig:overview}
\end{figure}

To answer the key question proposed in our preliminary study, we conduct a systematic research to improve and assess the dataset quality in terms of code smells. 
The research methodology overview is illustrated in Fig.~\ref{fig:overview}. 
We conduct our research with the following main steps: 
We first build an LLM-based code smell clean tool, namely \tool, to clean code smells within the existing benchmark dataset. 
\xt{Then we construct a test set of 50 repositories sourced from the \texttt{CodeSearchNet}-Python corpus for functional correctness testing.}
Following that, we fine-tuned the LLM with the cleaned version of the benchmark dataset and assessed whether the code smells of the LLM's generated code can be reduced. 
Thirdly, we choose two downstream tasks in this study, i.e., code generation and code search, to further verify the effectiveness of the cleaned version benchmark on specific software engineering tasks. 
Finally, we conducted a user study to obtain feedback from developers of whether code smells are cleaned correctly and comprehensively. 


\subsection{Research Questions}
Our study attempts to answer the following key research questions: 
\begin{enumerate}
    \item \textbf{RQ1: Can LLMs effectively clean code smells from training datasets?} 
    Although previous studies have applied LLMs to solve various code-related tasks, such as code generation~\cite{ni2023lever, Chen2024JumpCoderGB, svyatkovskiy2020intellicode}, test generation~\cite{ryan2024code, schafer2023empirical}, program repair~\cite{jin2023inferfix, wei2023copiloting, li2022generating}, their application in eliminating code smells has been rare or still unexplored. 
    Consequently, our initial focus is to assess how effectively LLMs can transform the original ``smelly'' training dataset into a smell-cleaned dataset. 

    \item \textbf{RQ2: Can LLMs effectively refactor and eliminate code smells while maintaining functional correctness consistent with the original code?}
    After performing code smell removal with LLMs, it is necessary to verify the functional correctness of the code before and after refactoring. So in this RQ, we strive to explore whether LLMs can maintain consistent code behavior while eliminating code smells.
    
    \item \textbf{RQ3: Can the smell-cleaned dataset alleviate the code smell problem within LLMs' generated code and thus enhance LLMs' output quality?}
    After cleaning code smells in the training dataset, we aim to investigate whether the curated smell-cleaned dataset can potentially lead LLM to generate higher-quality and more maintainable code with fewer code smells. 

    \item \textbf{RQ4: Can the smell-cleaned dataset enhance the model's performance in other downstream tasks?} 
    Cleaning code smells can enhance the code quality and maintainability, in this RQ, we seek to explore whether the smell-cleaned dataset can also benefit other downstream tasks. 
\end{enumerate}


\subsection{LLM-based Code Smell Cleaning Tool}
Inspired by the great potential of LLMs for code comprehension and code generation, we designed an LLM-based tool, named \tool, for cleaning code smells in the training dataset and curating a smell-cleaned dataset. 
The underlying approach of \tool is prompt engineering, i.e., using natural language to guide LLMs to complete specific tasks. 
Since LLMs are not designed for refactoring purposes, we leverage \textbf{prompt role designation}, \textbf{chain-of-thought reasoning}, and \textbf{few-shot learning} to harness the LLMs' knowledge for automated code smell refactoring.  

\begin{table}[htbp]
\setlength{\abovecaptionskip}{5pt}
\centering
\caption{The Example of Prompt Engineering}
\label{tab:interactive value predictor}
\begin{tabular}{p{3cm}|m{9cm}}
\toprule
\makecell[c]{Prompt Type} & \makecell[c]{Instantiation} \\
\bottomrule
\toprule
\multirow{3}{*}{Role Designation} 
& \textbf{Role:} \textit{You are an expert software engineer, with deep Python language expertise. Your task is to refactor the code to eliminate the code smell while the code function remains unchanged.}\\
& \textbf{Output Restriction:} \textit{Do not generate any explanation text nearby. No other symbols, comments, etc!}\\
\hline
\multirow{11}{*}{\makecell[c]{Chain-of-Thought\\Reasoning}}
& \textbf{Task:} \textit{Collapsible if statements mean that when two "if" statements are nested, we can improve the code's readability and reduce complexity by merging them. Please refactor it following the listing steps:}\\
& \textbf{Step 1: Understand the task requirement.} \textit{The goal is to merge two nested "if" or "elif" statements into a single line to remove Collapsible "if" Statements smell.}\\
& \textbf{Step 2: Analyze the content of two conditions.} \textit{Understand the functionality of both conditions.}\\
& \textbf{Step 3: Determine the conjunction.} \textit{Analyze the logical relationship between the two conditions.}\\
& \textbf{Step 4: Combine the conditions.} \textit{Use the appropriate conjunction and format to merge the conditions into one line while maintaining the original flow and logic.} \\
\hline
\multirow{7}{*}{\makecell[c]{Few-shot Learning\\Examples}} & \textbf{Given smell code:}\\
 & \quad \quad \textcolor{blue}{if} condition1:\\
 & \quad \quad \textcolor{blue}{\quad if} condition2:\\
 & \quad  \quad\quad \quad \# code\\
 & \textbf{Result Refactor Code:}\\
 & \quad \quad \textcolor{blue}{if} condition1 \textcolor{blue}{and} condition2:\\
 & \quad \quad \quad \# code\\
\bottomrule
\end{tabular}
\end{table}

\textbf{Prompt Role Designation.} 
In prompt engineering, role designation is a method where LLMs are designated a role for solving a specific task. 
Assigning a role to the LLM provides it with the problem context and leads to more accurate and relevant responses. 
In this study, we design the role of the LLM as ``\textit{an expert software engineer}''. 
After assigning the role, we clearly inform the LLM with our task as follows: ``\textit{Your task is to refactor code to eliminate code smell while keeping the code function unchanged.}'' 
The task description can elicit the programming knowledge of the LLM for performing this code refactoring task. 


\textbf{Chain-of-Thought Reasoning.} 
Chain-of-Thought (CoT) reasoning is an important strategy for prompt engineering~\cite{feng2024towards}. 
It enables the LLMs to split a complicated task into several relatively simple steps and generate a series of intermediate outputs that lead to a reasonable result. 
Our task of cleaning different code smells requires logical thinking to understand the code and a coherent series of intermediate steps to clean code smells. 
In order to construct a reasonable CoT, we invite two developers, with over 6 and 9 years of Python programming experience, to manually fix the 10 different code smells respectively and write down their core steps as their chain-of-thought reasoning steps. 
Afterwards, the first author discussed with the two developers to summarize the chain-of-thought reasoning steps for each code smell. 
We take the \textit{Collapsible If Statements} smell as an example to show the chain-of-thought reasoning process, the detailed CoT for other code smells can be found in our replication package. 
The smell-clean process is summarized as follows: (1) Step1: Understanding the task requirement; 
(2) Step2: Analyze the content of two conditions; 
(3) Step3: Determine the conjunction; (4) Combine the conditions. 
Overall, this step-by-step thinking guides the LLM to clean code smells in a manner similar to a developer.

\textbf{Few-shot Learning.} 
With the increasing ability of LLMs, in-context learning has been widely adopted as zero-shot learning and few-shot learning. 
Few-shot learning is utilized to augment the context with a few examples of desired inputs and outputs, which helps the model
elicit specific knowledge and abstractions needed to complete the task. 
Regarding the \textit{Collapsible If Statements} smell, we give an example of the code smell input (e.g., \texttt{if condition1: if condition2: \#code}) and the expected code smell output (e.g., \texttt{if condition1 and condition2: \#code}). 
With the help of these representative examples, the LLM can enhance its understanding of the target code smell and its effectiveness for refactoring.

\textbf{Tool Design.} 
The pipeline of our \tool works as follows: 
for a given code corpus, we first use SonarQube to automatically detect potential code smells within the code. 
After that, we leverage the LLM with each code smell's prompt to clean the code smell one by one according to the error location reported by error location reported by SonarQube. 
As a result, the original version of the training set will be cleaned by our tool into a cleaned version, denoted as smell-cleaned dataset. 

\textbf{Tool Implementation.} 
For LLMs, we chose the DeepSeek-Coder-V2 model (the latest version prior our experiment). 
DeepSeek-Coder has proven to have excellent abilities in coding tasks~\cite{zhu2024deepseek}, achieving a close and comparable performance with GPT-4o. 
Moreover, the DeepSeek-Coder model is rather cost-effective, compared with GPT-4o, the input token costs just 2.8\% and the output token costs just 1.9\% of the GPT-4o. 
Considering the benchmark dataset is relatively large, we chose the DeepSeek-Coder for this study due to its powerful ability and affordability. 
When using LLMs, the parameter of temperature is a pivotal setting that governs the randomness of the model's output. 
In this study, we set the temperature at 0 by a series of pilot experiments. 
According to the DeepSeek documentation, the \textit{max\_tokens} can reach 8,192. 
In this study, when handling code smells such as \textit{Long Parameter List}, we set the \textit{max\_tokens} to 8,192 to deal with the large number of tokens in the \textit{Long Parameter List}. 
For other cases, it was set to 2,048 to save time and resources.

\section{EVALUATION}

\subsection{RQ1. Effectiveness Evaluation}
In this research question, we aim to assess the effectiveness of our tool in cleaning the target code smells. 
To answer this RQ, we conduct an automatic evaluation and a manual analysis to estimate the \tool's clean power respectively.


\subsubsection{Automatical Evaluation.} 
Since all code smells in our research are identified by SonarQube, we first perform an automatic evaluation using SonarQube to verify whether the reported code smells have been effectively removed from within the smell-cleaned dataset. 
This involves detecting whether any of the previously reported code smells still exist or have been eliminated.


\noindent\textbf{Experiment Setup.} 
Specifically, based on the preliminary study results, the \texttt{CodeSearchNet}-Python dataset involves more than 200K code smells across ten different categories. 
\xt{In this RQ, we apply our \tool to perform targeted refactoring across the \texttt{CodeSearchNet}-Python dataset, specifically processing all of the more than 200K code smell instances identified by SonarQube while leaving smell-free components unchanged. 
The tool handles multiple coexisting smell types within individual Python files through in-place code modifications. 
As a result, this procedure converts the original dataset into a clean version dataset, denoted as smell-cleaned dataset.}
Following that, we run SonarQube on this smell-cleaned dataset once again to see if these existing code smells are addressed. 

\begin{table} 
\setlength{\abovecaptionskip}{0pt}
  \centering
  \caption{Automatically Detected Code Smell After Cleaning}
  \label{tab:process_smell}
  \begin{tabular}{lccc}
    \toprule
    Type &  \# Before Cleaning & \# After Cleaning & Cleaning Ratio(\%)\\
    \midrule
    \textit{Commented Code} &10,509 & \xt{735} & \xt{93.0} \\
    \textit{Naming Convention} & 133,323 & \xt{7,975} & \xt{94.0} \\
    \textit{Empty Nested Code Blocks}& 2,446 & \xt{226} & \xt{90.8}\\
    \textit{Collapsible if Statements}& 8,631 & \xt{185} & \xt{97.9}\\
    \textit{Long Parameter List} & 12,967 & \xt{5,502} & \xt{57.6} \\
   \textit{High Cognitive Complexity} & 33,926 & \xt{2,348} & \xt{93.1}\\
    \textit{Dead Code} & 458 & \xt{9} & \xt{98.0} \\
    \textit{Self-assigned Variables} & 458 & \xt{14} & \xt{96.9}\\
    \textit{Identical Expressions} & 267 & \xt{34} & \xt{87.3}\\
    \textit{Return and Yield} & 195 & \xt{47} & \xt{75.9}\\
    \midrule
    All & 203,180 & \xt{17,075} & \xt{91.6}\\
    \bottomrule
  \end{tabular}
\end{table}

\noindent\textbf{Experiment Result.}
Table~\ref{tab:process_smell} presents the code smell detection results of the original and the smell-cleaned dataset and the elimination ratio for each category of code smell. 
From the table, we can see that: 
(1) By comparing the number of detected code smells from SonarQube, \textbf{\xt{91.6\%} of code smells have been successfully eliminated from the original dataset, highlighting the impressive capability of our \tool in code smell cleaning}. 
\xt{
(2) According to the automatic evaluation results, our tool achieved a high clean ratio on \textit{Dead Code} smell, \textit{Collapsible If Statements} smell, \textit{Self-assigned Variables} smell, \textit{Naming Convention} smell, \textit{High Cognitive Complexity} smell, \textit{Commented Code} smell and \textit{Empty Nested Code Blocks} smell, over 90\% of these code smells are cleaned successfully. 
(3) Our \tool also shows relatively poor performance on several code smells (i.e., \textit{Long Parameter List} smell, \textit{Identical Expressions} smell and \textit{Return and Yield} smell). 
For example, only 57.6\% of the \textit{Long Parameter List} smells have been eliminated. 
}
This poor performance is likely due to the complicated data and control flow dependencies inherent in long parameter lists, as well as the challenges the LLM faces in processing and understanding lengthy code. 
The poor performance on these code smells also motivates us to conduct a more in-depth manual analysis.

\subsubsection{Manually Analysis}
\label{sec:manual_analysis}
The automatic evaluation only shows the overall number of code smells have been reduced, however, whether a code smell is really cleaned or not still needs manual validation. 
Moreover, the relatively poor performance on several code smells further motivates us to take a closer look at the cleaning process on each code smell. 
Therefore, besides the automatic evaluation, we conduct a manual analysis to further validate if the code smells are indeed cleaned. 

\noindent\textbf{Experiment Setup.} 
To conduct the human analysis, we randomly selected 387 code files (involving 1,272 Python functions) from the original \texttt{CodeSearchNet}-Python dataset. 
Such a sample guarantees a margin of error of ±5\% with a confidence level of 95\%.
Subsequently, the first two authors of this paper independently validated whether the code smells had been cleaned, they then meet and discuss the results until a consensus is reached. 

\begin{table} 
\setlength{\abovecaptionskip}{0pt}
  \centering
  \caption{Manual Validation Code Smell After Cleaning}
  \label{tab:manual_smell}
  \begin{tabular}{lccc}
    \toprule
    Type &  \# Before Cleaning & \# After Cleaning & Cleaning Ratio(\%)\\
    \midrule
    \textit{Commented Code} &68&\xt{6}& \xt{91.2}\\
    \textit{Naming Convention} &862&\xt{49}& \xt{94.3} \\
    \textit{Empty Nested Code Blocks}&13&\xt{2}&\xt{84.6} \\
    \textit{Collapsible if Statements}&54&\xt{4}& \xt{92.6}\\
    \textit{Long Parameter List} &13&\xt{7}& \xt{46.2} \\
    \textit{High Cognitive Complexity} &31&\xt{12}&\xt{61.3} \\
    \textit{Dead Code} &54&\xt{4}& \xt{92.6} \\
    \textit{Self-assigned Variables} &64&\xt{5}& \xt{92.2}\\
    \textit{Identical Expressions} &62&\xt{9}& \xt{85.5}\\
    \textit{Return and Yield} &51&\xt{6}& \xt{88.2}\\
    \midrule
    All & 1,272 & \xt{104}& \xt{91.8}\\
    \bottomrule
  \end{tabular}
\end{table}

\noindent\textbf{Experiment Result.} 
Table~\ref{tab:manual_smell} lists the manual validation results of the sampled code smells. 
The experimental results show that: 
(1) The overall performance of manual analysis is consistent with the automatic evaluation results, i.e., 91.9\% code smells are indeed cleaned by our tool, which further confirmed the effectiveness of \tool in removing code smells. 
\xt{(2) Compared with automatic evaluation, the clean ratio of the \textit{Return and Yield} smell is improved from 75.9\% to 88.2\%. We manually checked the differences and demonstrated an example for clarification in Sec.~\ref{sec:Strength}}. 
(3) We also found the clean ratio of \textit{High Cognitive Complexity} reduced from 93.1\% to 61.3\%. 
Based on our manual validation, we attribute the reduction to the calculation method of \textit{High Cognitive Complexity}, and we further discuss the detail in Sec.~\ref{sec:Weakness}. 
(4) After manual validation, the ability of our tool for cleaning \textit{Long Parameter List} is still limited. 
\xt{This is reasonable because how to correctly encapsulate the parameters within a dedicated dataclass structure or break a long method into smaller sub-methods while keeping its functionality is still an open research problem. We will also discuss in detail the difficulties that \tool encounters when dealing with this type of smell in Sec.~\ref{sec:Weakness}.}



\find{\textbf{Answer to RQ1:} Our \tool is highly effective for cleaning code smells from the existing benchmark dataset, more than 90\% code smells have been effectively cleaned and eliminated in the \texttt{CodeSearchNet}-Python benchmark dataset. 
}

\subsection{RQ2. Functional Correctness Verification} 
\label{sec:function_correctness_verify}
Although \tool can effectively remove code smells, we cannot be sure whether it ensures the consistency of code behavior. 
Therefore, in this RQ, we aim to verify the functional correctness of the code before and after refactoring based on the \texttt{CodeSearchNet}-Python dataset.

\xt{
\noindent\textbf{Test Set Preparation.} 
Since not all projects in \texttt{CodeSearchNet}-Python include test cases, we manually curated 50 testable projects from \texttt{CodeSearchNet}-Python to construct the evaluation dataset for RQ2. 
Specifically, to ensure broad coverage of different code smell types, we first ranked all projects according to the number of functions they contain. 
Then for each project, we manually assessed its testability by checking whether it includes test files or directories (e.g., test*.py) or uses testing frameworks (e.g., PyTest). 
A project was included in our evaluation dataset only if its tests can be successfully executed. 
Finally, we collected a subset of 50 testable projects from \texttt{CodeSearchNet}-Python.
Moreover, to verify the representativeness of our 50-project subset, we quantify and compare the prevalence of each smell type between the \texttt{CodeSearchNet}-Python corpus and our selected subset. Differential analysis revealed minimal divergence, with absolute percentage differences ranging from 0.01\% to 6.8\% across all smell categories (mean absolute difference=1.7\%). 
Specifically, the subset maintained comparable proportions for both prevalent smells (e.g., \textit{Naming Convention}: 72.4\% vs. 65.6\% in full corpus) and rare ones (e.g., \textit{Return and Yield}: 0.03\% vs. 0.1\% in full corpus). 
This close alignment demonstrates our sampled dataset preserved the natural distribution of smell occurrences while ensuring all target categories were adequately represented for comprehensive evaluation.
As a result, these selected 50 projects constitute our evaluation set for correctness verification. Each project is fully configured with prepared test suites, making them ready to run and analyze.
}

\begin{table} 
\small
\setlength{\abovecaptionskip}{0pt}
  \centering
  \caption{Code Smell Refactoring and Code Correctness Testing \xt{of 50 projects}}
  \label{tab:verify_functional_correctness}
  \xt{
  \begin{tabularx}{\textwidth} {cccccccc} 
    \toprule
    \multirow{2}{*}{Code Smell Type} & \multirow{2}{*}{Contri-} & \multicolumn{3}{c}{Code Smell Refactoring} & \multicolumn{3}{c}{Code Correctness Testing} \\
    \cmidrule(lr){3-5} \cmidrule(lr){6-8}
    & bution\% & \#Before & \#After & Cleaning(\%) & \#Before & \#After & Accuracy(\%) \\
    \midrule
    \textit{Naming Convention} & 74.37 & 11,579 & 62 & 99.5 & 11,579 & 10,578 & 91.4 \\
    \textit{High Cognitive Complexity} & 10.76 & 1,840 & 174 & 90.5 & 1,840 & 1,627 & 88.4 \\
    \textit{Commented Code} & 7.10 & 1,100 & 0 & 100.0 & 1,100 & 1,100 & 100.0 \\
    \textit{Long Parameter List} & 2.58 & 655 & 256 & 60.9 & 655 & 474 & 72.4 \\
    \textit{Collapsible if Statements} & 4.19 & 648 & 0 & 100.0 & 648 & 648 & 100.0 \\
    \textit{Empty Nested Code Blocks} & 0.60 & 109 & 16 & 85.3 & 109 & 107 & 98.2 \\
    \textit{Dead Code} & 0.24 & 37 & 0 & 100.0 & 37 & 37 & 100.0 \\
    \textit{Self-assigned Variables} & 0.08 & 13 & 0 & 100.0 & 13 & 13 & 100.0 \\
    \textit{Identical Expressions} & 0.05 & 8 & 0 & 100.0 & 8 & 8 & 100.0 \\
    \textit{Return and Yield} & 0.03 & 5 & 0 & 100.0 & 5 & 5 & 100.0 \\
    \midrule
    All & 100.0 & 15,994 & 508 & 96.8 & 15,994 & 14,597 & 91.3 \\
    \bottomrule                                
  \end{tabularx}
  }                                  
\end{table}


\noindent\textbf{Experiment Setup.} 
For each project, we initially ran test cases on the original code before refactoring, and the results confirmed that all tests passed successfully. 
Subsequently, we ran SonarQube to detect code smells and used \tool to clean the detected smells, we ensured the correctness of the refactored code by rerunning the test cases.
Then we further calculated the test accuracy by the number of tests passed before and after the refactoring.
\xt{
Finally, to quantitatively assess the contribution of different refactoring types to performance improvements, we calculated each type's contribution rate as its proportion of the total refactoring instances.
It is worth mentioning that, to ensure backward compatibility with existing test suites while addressing the \textit{Long Parameter List} smell, the proposed refactoring strategy employs a conservative two-phase approach: first, maintaining the original function signature as an adapter layer to preserve test accessibility, while simultaneously introducing a new implementation that encapsulates the parameters within a dedicated dataclass structure. 
This methodology leverages the adapter design pattern to decouple interface stability from internal implementation improvements, where the legacy function serves as a thin wrapper that instantiates the parameter dataclass and delegates core logic to the refined function. 
}

\xt{
\noindent\textbf{Experiment Result.}
Table~\ref{tab:verify_functional_correctness} presents the results of code smell elimination and functional correctness testing of 50 projects.
The experimental results show that SonarQube detected 15,994 code smells from the above projects, and \tool \textbf{successfully removed 96.8\%} of them (15,486 out of 15,994). 
By running unit tests, we found that \textbf{91.3\%} (14,597 out of 15,994) of the refactored code maintained the same functionality as the original one. 
From perspective of different code smell types, several points stand out:

\noindent\faThumbsUp~\textbf{Perfect Refactoring Cases}: 
From the table, we can see that \tool shows perfect (i.e., 100\%) effectiveness and correctness for six smells: \textit{Commented Code}, \textit{Collapsible if Statements}, \textit{Dead Code}, \textit{Self-assigned Variables}, \textit{Identical Expressions} and \textit{Return and Yield}. 
These code smells are relatively straightforward to detect and refactor by syntactic patterns or pre-defined rules, the perfect performance in these cases highlights \tool’s strong capability in handling structurally simple yet common syntactic code issues.
    
\noindent\faThumbsUp~\textbf{High-Performance Cases}: 
\tool is highly effective for cleaning following three code smells: \textit{Naming Convention} (99.5\% effectiveness and 91.4\% correctness), \textit{High Cognitive Complexity} (90.5\% effectiveness and 88.4\% correctness) and \textit{Empty Nested Code Blocks} (85.3\% effectiveness and 98.2\% correctness). 
These three types of code smells requires a deeper understanding of code semantics and developer’s intent, which goes beyond the surface-level pattern matching. 
For instance, addressing \textit{High Cognitive Complexity} requires the model to recognize logical structures, abstract control flow, and restructure the code while preserving its behavior. 
The success of \tool on these cases demonstrate strong semantic comprehension capabilities, which are able to interpret code context, reason about its functionality, and generate meaningful edits. 
This also highlights the superiority of the LLM-based method in handling non-trivial code quality issues over the traditional rule-based refactoring methods. 

\noindent\faThumbsDown~\textbf{Challenging Cases}: 
The \textit{Long Parameter List} smell demonstrates the lowest refactoring effectiveness among all ten categories, achieving only a 60.9\% cleaning rate with 72.4\% correctness, this relatively poor performance may be attributed to following reasons: 
First, refactoring \textit{Long Parameter List} often requires a broader understanding of program context and project-specific knowledge. For example, grouping related parameters into new data structure or removing redundant parameters are context-sensitive, this level of reasoning is still challenging for LLMs, especially when the necessary context spans across multiple files or functions. 
Second, resolving \textit{Long Parameter List} often involves coordinated changes across multiple functions or modules (e.g., updating function calls and dependent code), increasing the complexity and likelihood of fixing the \textit{Long Parameter List} smell. 
These findings reveal the limitation of \tool in handling smells that require global reasoning, suggesting future improvements through extending context-window and employing multi-agents collaboration across multiple functions and files. 

Also, table~\ref{tab:verify_functional_correctness} shows the contribution rates of different code smell categories to overall refactoring efforts. 
Notably, the contribution analysis reveals a highly skewed distribution, with \textit{Naming Convention} smells dominating (74.39\% of all refactorings) while the remaining nine categories collectively account for merely 25.61\%. 
But this contribution pattern strongly aligns with the natural occurrence frequency of these smell types in the test set (Pearson's r = 1.00, p < 0.001) and in the \texttt{CodeSearchNet}-Python (Pearson's r = 0.992, p < 0.001).

\find{\textbf{Answer to RQ2:} Our \tool can effectively fix most of the code smells (i.e., 96.8\%) while maintaining the same functionality (i.e., 91.3\%).}
}

\subsection{RQ3. LLMs' Generated Code Quality Improvement}
In RQ1, we have converted the \texttt{CodeSearchNet}-Python benchmark into a smell-cleaned dataset. 
Since the data quality has a direct impact on models' performance~\cite{shi2022we, croft2023data, gao2024learning, cote2024data}, in this RQ, we aim to investigate whether the smell-cleaned dataset can improve LLMs' generated code quality. 

\noindent\textbf{Experiment Setup.}
For a fair comparison, we used the same test set from the preliminary study, which includes 1,000 randomly sampled descriptions, while the remaining data was employed as the training dataset. 
\xt{To confirm our hypothesis, we fine-tuned three different sizes of LLM: DeepSeek-Coder-V2-Lite-Instruct (16B), Qwen2.5-Coder-7B-Instruct (7B),
and DeepSeek-Coder-6.7B-Instruct (6.7B), which suffered from generating code smells, with two datasets, the original version of \texttt{CodeSearchNet}-Python dataset and our curated smell-cleaned version, obtaining six fine-tuned models.
We then leverage nine LLMs, i.e., six fine-tuned models and three base models, on the 1,000 samples, generating target code based on their descriptions. 
Finally, we apply SonarQube to detect potential code smells within nine LLMs' generated code respectively.}
\xzp{This experiment is conducted with four NVIDIA A800 SXM4 80GB GPUs.
To avoid potential biases from sample selection~\cite{cortes2008sample} or parameter adaptation~\cite{stanovov2021biased}, we utilize full fine-tuning technology to precisely ascertain the impact attributable to the clean/smell dataset. 
We configured the model fine-tuning following the implementation provided in their project documentations~\cite{DeepSeek-Coder_Project} and adopted the recommended hyperparameter settings. 
We set the training epoch to 2 to save training time and computation resources~\cite{guo2024deepseek,bai2023qwen}. 
The fine-tuning process took approximately 72 hours for DeepSeek-V1 and Qwen-Coder, and 144 hours for DeepSeek-V2.}

\begin{table} 
\small
\setlength{\abovecaptionskip}{0pt}
  \centering
  \caption{\xt{Detected Code Smells in LLM Generated Code}}
  \label{tab:gene_code}
  \xt{
  \begin{tabular}{cccccccccc}
    \toprule     
    \multirow{2}{*}{Type} & \multicolumn{3}{c}{DeepSeek-V1} & \multicolumn{3}{c}{DeepSeek-V2} & \multicolumn{3}{c}{Qwen-Coder} \\
    \cmidrule(lr){2-4} \cmidrule(lr){5-7} \cmidrule(lr){8-10}
     & base & \#origin & \#clean & base & \#origin & \#clean& base & \#origin & \#clean\\
    \midrule
    \textit{High Cognitive Complexity}&12&35&11&6&5&4&11&10&6\\
    \textit{Long Parameter List}&9&10&2&8&10&7&10&9&1\\
    \textit{Naming Convention}&116&84&39&105&124&16&95&99&15\\
    \textit{Collapsible if Statements}&11&21&9&13&14&7&9&9&4\\
    \textit{Commented Code}&0&0&0&0&2&0&0&5&0\\
    \textit{Empty Nested Code Blocks}&3&1&1&2&1&0&1&12&0\\
    \textit{Dead Code}&0&0&0&0&0&0&0&2&0\\
    \textit{Self-assigned Variables}&1&8&0&3&11&0&0&1&0\\
    \textit{Return and Yield}&0&0&0&0&0&0&0&7&0\\
    \midrule
    All & 152 & 159 & 62 & 137 & 167 & 34 & 126 & 154 & 26 \\
    \bottomrule
  \end{tabular}
  }
\end{table}

\noindent\textbf{Experiment Result.} 
The number of code smells detected by SonarQube in the code generated by each model is illustrated in Table~\ref{tab:gene_code}. 
From the table, several points stand out: 
\xt{(1) We observed a significant decrease in code smells in codes generated by all three different models fine-tuned on the smell-cleaned dataset.
In other words, \textbf{fine-tuning the LLM with the smell-cleaned dataset can significantly improve the LLM's generated code quality in terms of code smells. }
Specifically, fine-tuning with the smell-cleaned dataset reduces code smell issues by 61.0\% for DeepSeek-V1, 79.6\% for DeepSeek-V2, and 83.1\% for Qwen-Coder compared to using the original dataset - markedly higher than the 59.2\%, 75.2\%, and 79.4\% reductions achieved over their respective base models without fine-tuning - suggesting that the smell-cleaned training dataset can be used to improve LLMs' generated code quality.
These results highlight that data quality is pivotal: proper dataset cleaning both eliminates the harmful effects of noisy training data and enables consistent enhancements to LLM-generated code quality.
(2) By comparing the performance of models fine-tuned on the origin dataset with corresponding base versions, the code smell problems are becoming even more serious, which further confirms the original \texttt{CodeSearchNet}-Python has quality issues with code smells and the training set quality can affect the LLM's output quality. }


\find{\textbf{Answer to RQ3:} The smell-cleaned dataset can be used to fine-tune the LLM to generate higher-quality and more maintainable code with fewer code smells.}

\subsection{RQ4. Downstream Tasks Performance}

Code smells are directly related to code quality and maintainability, however, code smells are often considered to have limited effects on code functionality. 
In this RQ, we aim to investigate whether addressing these code smells can have a positive influence in other downstream tasks. 
\xt{To verify our hypothesis, we conduct an experiment on two widely used software engineering tasks, i.e., code completion and code search.} 
\xzp{Specifically, to ensure generalizability of our conclusions, we evaluate model performance before and after applying smell-cleaning using three LLMs of different sizes: DeepSeek-Coder-V2-Lite-Instruct (16B), Qwen2.5-Coder-7B-Instruct (7B), and DeepSeek-Coder-6.7B-Instruct (6.7B).}

\begin{figure} 
    \centering
    \includegraphics[width=0.78\textwidth]{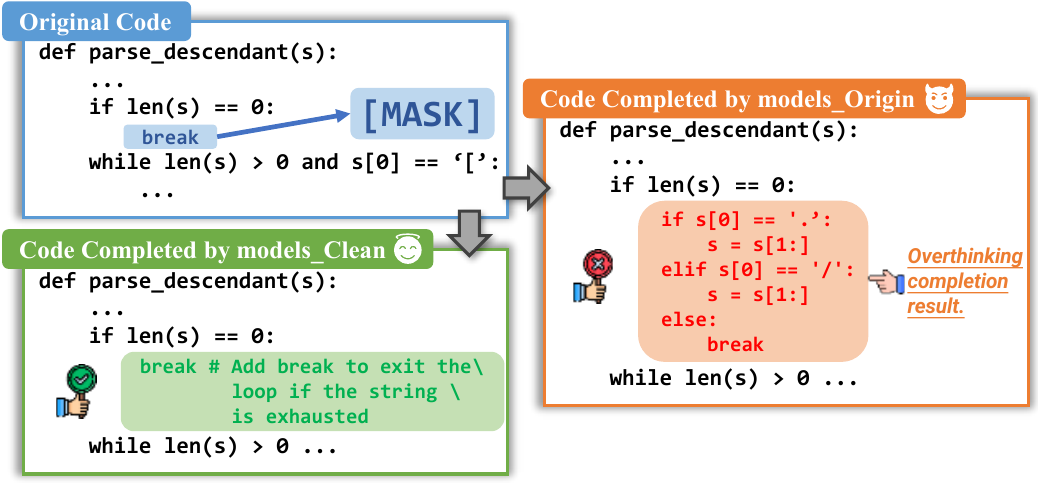}
    \caption{Successful Example of Code Completion Performance Improvement}
    \label{fig:code_completion_example}
\end{figure}

\xt{
\subsubsection{Code Completion Task.} 
Code completion is an important task in software development and has attracted attention from both academic and industry practitioners, especially when LLMs come out~\cite{Chen2024JumpCoderGB, mu2024clarifygpt, bai2023qwen}. 
Specifically, the code completion task is formulated as predicting the missing or subsequent code tokens given preceding code context (i.e., prefix completion) or surrounding context (i.e., fill-in-the-middle completion). 
In this RQ, we aim to investigate whether fine-tuning LLMs on our smell-cleaned dataset can enhance their performance on the code completion task. 


\noindent\textbf{Experiment Setup.}
We employ the standardized test set from RQ2 to comprehensively evaluate all models' performance on the code completion task.
To construct our evaluation benchmark, we randomly sampled 1,000 functions from the standardized test set. 
For each function, we randomly masked a single code line, ensuring the masked code line contains meaningful logic rather than trivial syntax (e.g., braces or blank lines). 
The candidate models are then utilized to predict the masked content based on the remaining code context. 
In this RQ, we conducted code completion experiments using three LLMs (i.e., DeepSeek-V1, DeepSeek-V2, and Qwen-Coder) under three conditions: (1) the original base models without fine-tuning, (2) models fine-tuned on the original dataset containing code smells, and (3) models fine-tuned on our curated smell-cleaned dataset.
The fine-tuning process and setup are identical to those described in RQ3.
Notably, the test set is removed from the smell-cleaned dataset for fine-tuning.
The \textit{Pass@1} metric is used to assess the precision of the completed code, measuring whether the completed code passes the original test. 
This metric reflects the model's accuracy for generating semantically correct code.

\begin{table} 
\setlength{\abovecaptionskip}{5pt}
  \centering
  \caption{\xt{Code Completion Performance of Different Models fine-tuned on Different Datasets}}
  \label{tab:code_completion}
\xt{
\begin{tabular}{lccc}
    \toprule
    \multirow{2}{*}{Models} & \multicolumn{1}{c}{DeepSeek-V1} & \multicolumn{1}{c}{DeepSeek-V2} & \multicolumn{1}{c}{Qwen-Coder} \\
     & Pass@1 & Pass@1 & Pass@1 \\
    \midrule
    Base Model & 0.750 & 0.730 & 0.816 \\
    Fine-tuned on Origin Dataset & 0.731 & 0.664 & 0.787 \\
    Fine-tuned on Clean Dataset & \textbf{0.789} & \textbf{0.742} & \textbf{0.883} \\
    \bottomrule
\end{tabular}
}
\end{table}

\noindent\textbf{Experiment Result.} 
Table~\ref{tab:code_completion} illustrated the code completion performance of base model and fine-tuned LLMs. 
According to the experimental results, several points stand out: 
(1) \textbf{LLMs fine-tuned with the smell-cleaned dataset outperform both base LLMs and LLMs fine-tuned with the original dataset}. 
Compared to base versions, the models fine-tuned on our smell-cleaned dataset showed significant improvements of 5.2\%, 1.6\%, and 8.2\% for DeepSeek-V1, DeepSeek-V2, and Qwen-Coder, respectively. 
Additionally, when compared to models fine-tuned on the original dataset, the smell-cleaned version further enhanced performance by 7.9\%, 11.7\% and 12.2\%.
The improvements across all models support our hypothesis that high-quality datasets not only enable LLMs to generate higher-quality code but also enhance the overall correctness of the generated code. 
(2) \textbf{After fine-tuning with the original \texttt{CodeSearchNet}-Python dataset, all models exhibited reduced accuracy compared to their base counterparts.} 
This highlights potential noise and data quality issues in the original \texttt{CodeSearchNet}-Python dataset. 
The observed declined performance (e.g., 2.5\% - 9.0\%) further confirms the critical role of data quality in model construction and refinement.
(3) Surprisingly, the smaller 7B-parameter DeepSeek-V1 model consistently outperformed its larger 14B-parameter counterpart (DeepSeek-V2) across three conditions. 
This may be because the smaller models, with less capacity, tend to focus on the most consistent and effective patterns, which can help in noisy-data settings. 

To better illustrate the benefits of our smell-cleaned dataset, we analyze a representative example from the benchmark where the model fine-tuned on smell-cleaned data correctly predicted the code completion, while the original-dataset-fine-tuned model failed. 
As shown in Fig.~\ref{fig:code_completion_example}, given an input code snippet with a masked line (i.e., \texttt{if len(s) == 0: [MASK]}), the model fine-tuned on the smell-cleaned dataset correctly predicts the original control flow (i.e., \texttt{break}), also adding a clarifying comment explaining the control flow. 
In contrast, the model fine-tuned on the original noisy dataset introduces a nested conditional (i.e., \texttt{if s[0] == '.'} and \texttt{elif s[0] == '/'}), which is both redundant (as is already handled later) and risky (raising an IndexError when \texttt{s} is empty). 
Such undesirable completions reflect that the \textit{High Cognitive Complexity} constructs prevalent in the original dataset — characterized by redundant nesting, excessive branching, and inflated control-flow depth — can negatively mislead model learning and hinder its ability to generate concise and correct code. 
}

\subsubsection{Code Search Task.}
In modern software development, code search is one of the most frequent activities~\cite{liu2021opportunities, gao2023know}. 
Code search refers to the retrieval of relevant code snippets from a code base, according to the intent of a search query. 
Recent studies~\cite{zeng2023degraphcs, sun2022code} tried to use deep learning techniques and/or LLMs to perform the code search task. 
The core idea is to encode the search query and the code snippet into high-dimensional embedding vectors, and then calculate the similarity score between the vectors to estimate their matching scores. 
After cleaning the obsolete information (e.g., \textit{Commented Code} smell, \textit{Dead Code} smell) and imprecise semantics (e.g., \textit{Naming Convention} smell) from the code, we assume it is possible for a code search model to achieve better performance on the code search task by using our curated smell-cleaned dataset. 

\noindent\textbf{Experiment Setup.} 
\xt{ 
In this RQ, we also employ DeepSeek-V1, DeepSeek-V2, and Qwen-Coder as base models for code search, evaluating them on both the original and smell-cleaned versions of the \texttt{CodeSearchNet}-Python dataset.
}
After filtering out comments that are either too short (e.g., less than 10 words) or too long (e.g., more than 200 words) or contain URLs, we randomly sampled 1,000 comment-code pairs as our testing set. 
The comment is regarded as the search query and its paired code is the target code to retrieve. 
For each comment, we randomly selected 100 code snippets (including the target code) to construct a code candidate pool as the codebase. 
\xt{We then use the DeepSeek-V1, DeepSeek-V2, and Qwen-Coder to encode the search query} and each candidate code into vector representations and calculate the matching score between two vectors based on their cosine similarities. 
We adopted the \textit{MRR} and \textit{NDCG} metric to evaluate the model performance, which measures how often the target code snippet can be successfully retrieved among other code snippet candidates. 

\begin{table} 
\setlength{\abovecaptionskip}{5pt}
  \centering
  \caption{\xt{Code Search Performance with Different Datasets}}
  \label{tab:code_search}
\xt{
\begin{tabular}{lcccccc}
    \toprule
    \multirow{2}{*}{Dataset} & \multicolumn{2}{c}{DeepSeek-V1} & \multicolumn{2}{c}{DeepSeek-V2} & \multicolumn{2}{c}{Qwen-Coder} \\
    \cmidrule(lr){2-3} \cmidrule(lr){4-5} \cmidrule(lr){6-7}
     & \textit{MRR} & \textit{NDCG} & \textit{MRR} & \textit{NDCG} & \textit{MRR} & \textit{NDCG} \\
    \midrule
    Original Dataset & 0.564 & 0.604 & 0.752 & 0.799 & 0.631 & 0.672 \\
    Smell-Cleaned Dataset & \textbf{0.587} & \textbf{0.620} & \textbf{0.757} & \textbf{0.802} & \textbf{0.656} & \textbf{0.701} \\
    \bottomrule
\end{tabular}
}
\end{table}

\begin{figure} 
    \centering
    \includegraphics[width=0.78\textwidth]{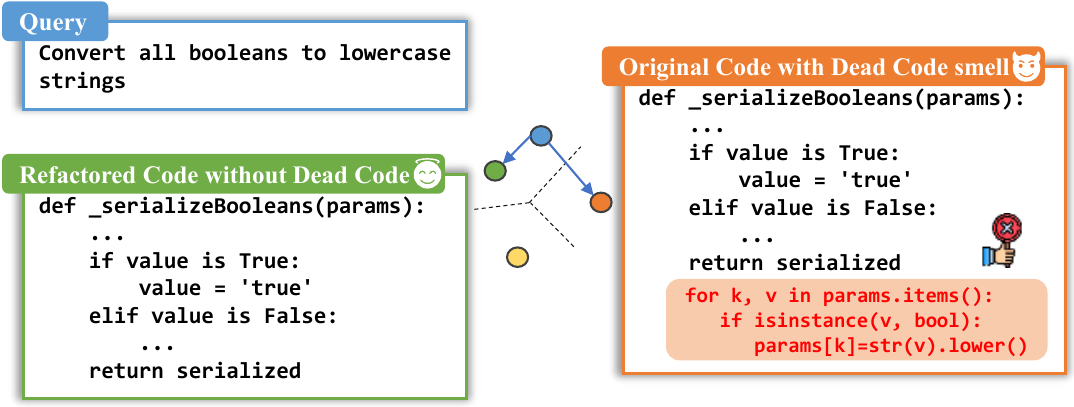}
    \caption{Successful Example of Code Search Performance Improvement}
    \label{fig:code_search_example}
\end{figure}

\noindent\textbf{Experiment Result.} 
Table~\ref{tab:code_search} shows the code search performance with respect to the original dataset and the smell-cleaned dataset. 
\xt{
We found that: (1) All evaluated models (i.e., DeepSeek-V1, DeepSeek-V2, and Qwen-Coder) consistently demonstrated superior code search performance on the smell-cleaned dataset. 
Specifically, when evaluated on the clean dataset compared to the original ``smelly'' dataset, all models show consistent improvements: DeepSeek-V1 achieves performance gains of 4.1\% (\textit{MRR}) and 2.6\% (\textit{NDCG}), DeepSeek-V2 improves by 0.7\% (\textit{MRR}) and 0.4\% (\textit{NDCG}), and Qwen-Coder demonstrates increases of 4.0\% (\textit{MRR}) and 4.3\% (\textit{NDCG}).
The improvements across all models support our assumption that high-quality datasets not only enable LLMs to generate higher-quality code but also enhance the semantic consistency between query and code, thereby improving search performance. 
(2) This performance boost is visually demonstrated with an example in Fig.~\ref{fig:code_search_example}, where the query ``\textit{Convert all booleans to lowercase strings}'' correctly retrieves the refactored version (similarity score: 0.81, rank@1) without \textit{Dead Code} smell.
In contrast, the original code version (similarity: 0.781) ranks at position 3, its performance degraded by semantic noise from dead code - particularly the unreachable loop following the \texttt{return} statement. 
Such noise can introduce irrelevant lexical patterns, distorting the LLM embeddings and negatively impacting the LLM's code comprehension. 
Overall, the experimental results confirm that eliminating code smells improves semantic consistency between queries and code, reducing noise-induced ranking drops — even for strong models like DeepSeek-V2. 
}

\find{\textbf{Answer to RQ4:} Smell-cleaned dataset can be used to enhance the model's performance in downstream tasks, such as code completion and code search.}
\xt{
\section{Ablation Study} 
To better understand the impact of different components in prompts(i.e., Role Designation, Chain-of-Thought Reasoning, and Few-shot Learning Examples) on overall refactoring performance, we conduct an ablation study on \tool. 
This analysis aims to: (1) quantify the relative importance of each design element in our prompting strategy, and (2) identify potential synergistic effects between components. 

\noindent\textbf{Experiment Setup.} 
We design the ablation study by systematically constructing four prompt configurations. 
Specifically, we compare the following variants on the code smell elimination task: (1) Role Designation only (denoted as Role-only), (2) Role + Few-shot Learning (denoted as Role+Few-shot), (3) Role + Chain-of-Thought (denoted as Role+CoT), and (4) the full combination (Role + CoT + Few-shot) to examine synergistic effects. 
Each variant is applied to identical code refactoring tasks, with the primary metric being code smell elimination rate. 
Such a design allows us to isolate the impact of each prompt component, validate their individual effects, and ultimately identify the optimal prompt engineering strategy for maximizing refactoring efficacy.

\begin{table} 
\setlength{\abovecaptionskip}{0pt}
  \centering
  \caption{\xt{Ablation Study with Different Prompt Configurations}}
  \label{tab:ablation_of_prompts}
  \xt{
  \begin{tabular}{lccc}
    \hline
    prompts           & \# Before & \# After & Cleaning(\%) \\ \hline
    Role-only              & 15,994    & 1,081    & 93.2         \\
    Role+Few-shot     & 15,994    & 1,050    & 93.4         \\
    Role+CoT          & 15,994    & 641     & 96.0         \\
    Role+CoT+Few-shot & 15,994    & 508     & 96.8         \\ \hline
  \end{tabular}
  }
\end{table}

\begin{figure} 
    \centering
    \includegraphics[width=0.56\textwidth]{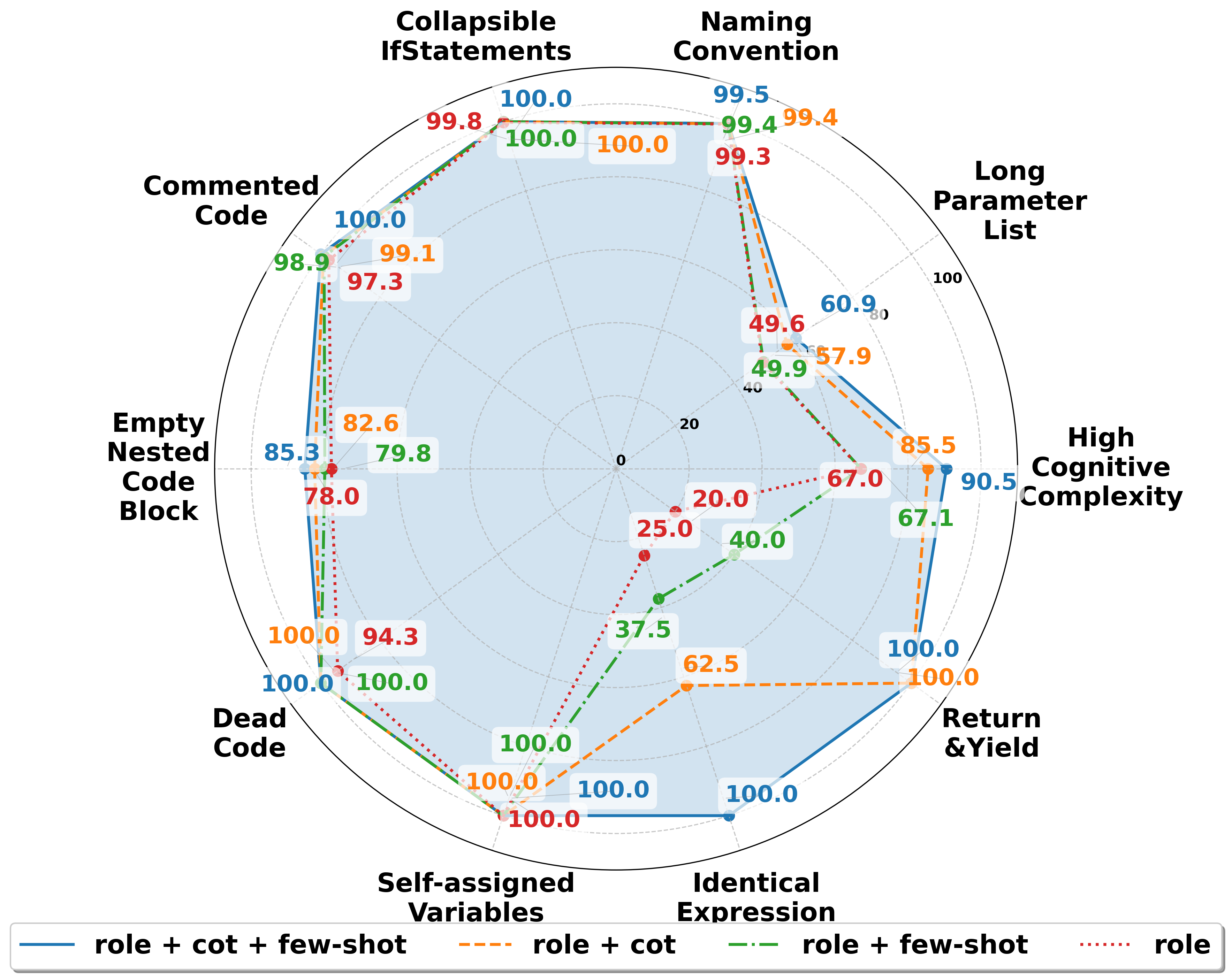}
    \caption{Comparative Performance on 10 Smell Types Under Different Prompt Settings}
    \label{fig:radar_of_prompt_ablation}
\end{figure}

\noindent\textbf{Experiment Result.}
The results are shown in Table~\ref{tab:ablation_of_prompts} and the accompanying radar chart (Fig.~\ref{fig:radar_of_prompt_ablation}) which comparatively visualizes all 10 code smell categories' elimination rates across the four prompt configurations, revealing distinct performance patterns for each smell type.
As presented in Table~\ref{tab:ablation_of_prompts} and Fig.~\ref{fig:radar_of_prompt_ablation}, our study yields five significant insights into the individual and combined effects of each \tool component: 
\textbf{(1)} The Role-only baseline achieves a 93.2\% code smell elimination rate, demonstrating the strong performance of the foundation DeepSeek-Coder base model in code refactoring tasks. 
\textbf{(2)} The integration of Chain-of-Thought Reasoning improves performance to 96.0\%, highlighting its critical role in guiding structured, step-by-step reasoning for more accurate refactoring decisions. This improvement is consistently observed across individual categories of code smells, further validating the generalizability of the approach in diverse refactoring scenarios.
\textbf{(3)} In contrast, the addition of Few-shot Learning results in only a marginal improvement (93.4\%), suggesting that while examples provide some contextual guidance, their standalone impact is limited compared to explicit reasoning augmentation.
The performance gain of SmellCC over Role + CoT alone confirms that Few-shot Learning provides complementary benefits when combined with structured reasoning, particularly in edge cases where exemplars reinforce CoT-generated rationales.
\textbf{(4)} The full framework (Role + CoT + Few-shot) achieves the highest performance (96.8\%), surpassing all partial configurations. This optimal result stems from synergistic interactions between the three techniques: (a) Role designation ensures task-specific focus, (b) CoT enables systematic decomposition of refactoring steps, and (c) Few-shot Learning supplements reasoning with contextualized examples, reducing ambiguity in complex cases.
\textbf{(5)} The SmellCC framework's 0.8\% gain over Role + CoT alone confirms that Few-shot Learning provides complementary benefits when combined with structured reasoning, particularly in edge cases where exemplars reinforce CoT-generated rationales. 
Overall, these findings validate our hypothesis that integrating role-based prompting, explicit reasoning, and example-driven learning maximizes refactoring efficacy by addressing both logical and contextual dimensions of code improvement.
}

\section{Discussion}

\subsection{Case Study}
To better understand the strengths and weaknesses of \tool, we conducted a case study based on the previous experimental results to analyze why our tool works and fails.  

\begin{figure} 
    \centering
    \subfloat[Example of Naming Convention]{%
       \includegraphics[width=0.45\textwidth]{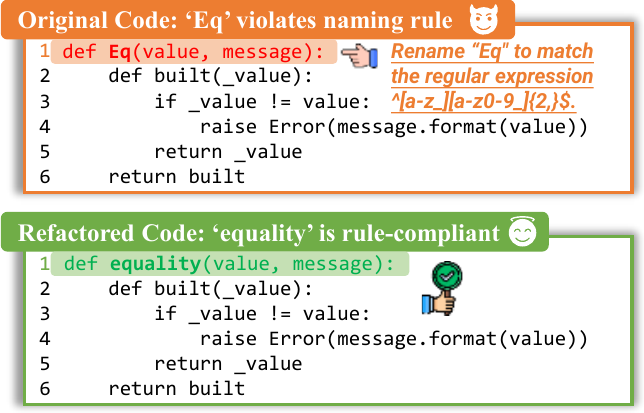}\label{fig:name convention}} 
    \qquad
    \subfloat[Example of Self-assigned Variables]{%
       \includegraphics[width=0.45\textwidth]{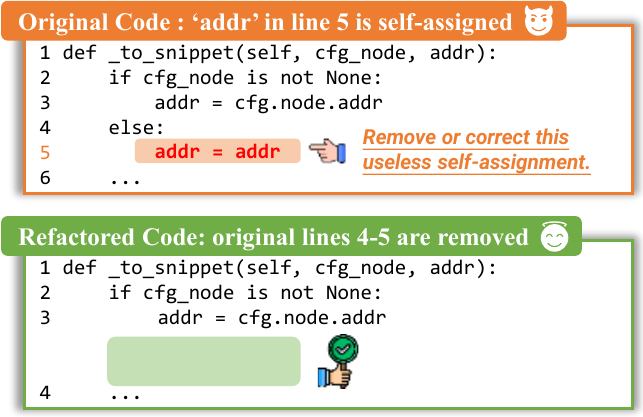}\label{fig:self-assigned variables}} 
    \caption{Successful Refactored Cases}
    \label{fig:successful_case}
\end{figure}


\subsubsection{Strengths Analysis.}
\label{sec:Strength}
Fig.~\ref{fig:successful_case} illustrates two instances which have been confirmed that cleaned the code smell issues while maintaining the same functionality. 
According to these two confirmed examples, we conclude two strengths for our approach to code refactoring: 
(1) \textbf{\tool demonstrates a strong ability in code comprehension.}
Specifically, Fig.~\ref{fig:name convention} illustrates the original and refactored versions of a \textit{Naming Convention} smell. 
\xt{SonarQube flags the function name \texttt{Eq} as violating naming conventions, suggesting it should be renamed to match the regular expression \texttt{\^{}[a-z\_][a-z0-9\_]\{2,\}\$}.
Then \tool replaces the ambiguous function name \texttt{Eq} with the more comprehensible and rule-compliant, i.e., \texttt{equality}. 
}
By refactoring this \textit{Naming Convention} smell, \tool improves semantic consistency between the function name and its function body, thereby enhancing both code readability and maintainability. 
(2) \textbf{\tool presents effective code editing capabilities.}
Fig.~\ref{fig:self-assigned variables} illustrates a \textit{Self-assigned Variables} smell, along with its corresponding refactored code. The statement ``\texttt{else: addr = addr}'' is redundant and does not alter the code’s semantics but negatively impacts its performance and increases cognitive complexity. 
\xt{According to SonarQube’s analysis, this self-assignment should be removed or corrected to improve code quality.
Encouragingly, \tool effectively removed the redundant else statement, ensuring the syntax correctness of the refactored code.
Compared with the traditional rule-based code refactoring method, we attribute \tool’s strengths to the following aspects: (i) Code semantic understanding: Different from traditional rule-based methods, \tool is based on LLMs, which has great ability for understanding code context and semantics; (ii) Generalizability: the rule-based methods are too complex and specific, which can hardly generalize to new added types of code smells. On the contrary, \tool is based on prompt engineering, which is extremely lightweight and flexible; (iii) We further leverage few-shot learning and CoT reasoning prompting techniques to elicit human knowledge and logical reasoning from LLMs to accurately identify each type of code smells and clean code smells in a manner similar to a skilled developer.


\begin{figure}
    \centering
    \includegraphics[width=0.45\textwidth]{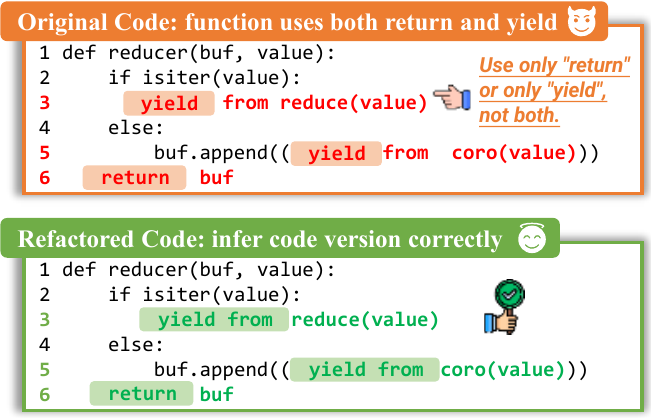} 
    \caption{False Positive Case detected by SonarQube}
    \label{fig:false_positive_case}
\end{figure}

Beyond the successfully refactored cases, SmellCC also shows strong infer capability.
Fig.~\ref{fig:false_positive_case} presents a false postive example of \textit{Return and Yield} smell reported by SonarQube. 
Before Python 3.3, using both \texttt{return} and \texttt{yield} in the same function would result in a \texttt{SyntaxError}. 
Based on this rule, SonarQube wrongly identifies the original code as smelly code with the diagnostic message: ``Use only \texttt{return} or only \texttt{yield}, not both''.
However, the example code targets a version later than Python 3.3, as indicated by the use of \texttt{yield from}, a feature introduced in Python 3.3. 
\tool accurately infers the current Python version and identifies it as smell-clean, thus avoiding any refactoring. 

\find{\textbf{Summary 1:} \tool \textbf{fundamentally outperforms rule-based tools by demonstrating exceptional code editing and human-like reasoning capabilities}—semantically understanding code context, inferring implicit constraints (e.g., version compatibility), and making intent-aware preservation decisions that traditional rule-based methods cannot.}
}


\begin{figure}
    \centering
    \includegraphics[width=0.95\textwidth]{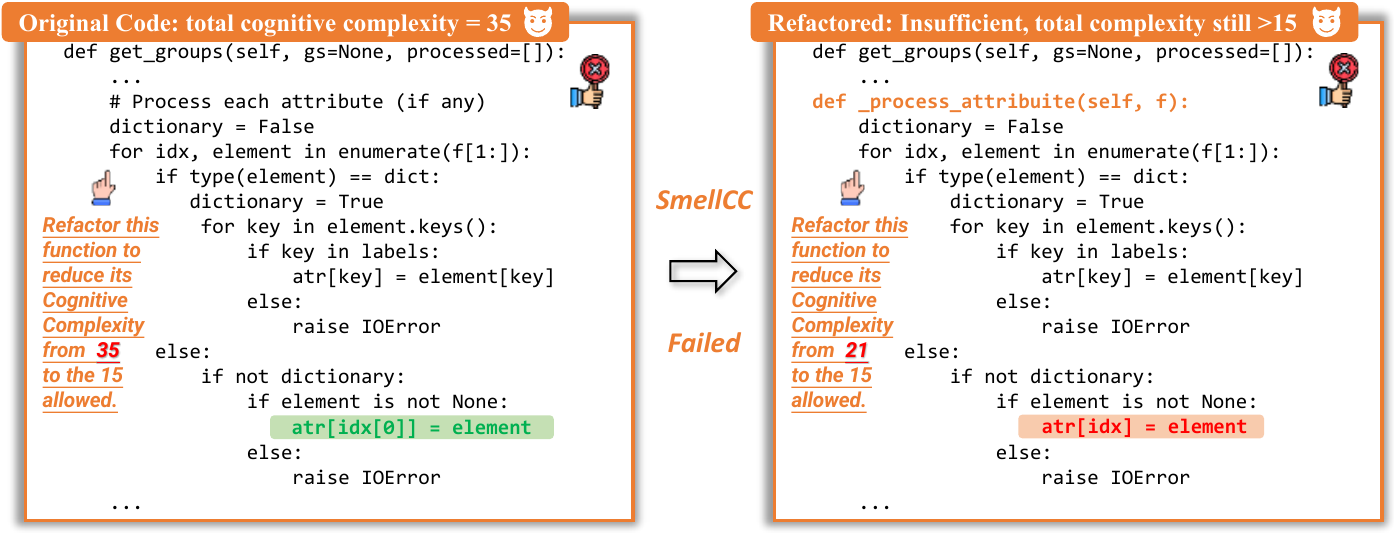} 
    \caption{A Failed High Cognitive Complexity Case}
    \label{fig:fail_case}
\end{figure}

\subsubsection{Weakness Analysis.}
\label{sec:Weakness}
Our approach performs relatively poorly on two specific types of code smells, i.e., 
\textit{High Cognitive Complexity} smell and \textit{Long Parameter List} smell. 
We manually reviewed failed cases of these two code smells to further analyze the weakness of our approach. 
Fig.~\ref{fig:fail_case} presents a \textit{High Cognitive Complexity} smell that our tool failed to refactor.
\xt{The original code is a function with a cognitive complexity of 35. 
The cognitive complexity of a function is calculated by assessing the nested control flow structures, logical operators, and breaks in linear execution, with weightings for nesting depth and structural complexity.
Therefore, according to SonarQube's quality standards, any function scoring above 15 in cognitive complexity is classified as having the \textit{High Cognitive Complexity} smell, indicating poor maintainability. 
Although the refactored version reduces the complexity by 14 through method decomposition, the refactored code remains above the threshold - indicating insufficient refactoring.
Through further investigation, we conclude the following three weaknesses of our approach: 
(1) \textbf{The LLM's incomplete refactoring.} We found that incomplete refactoring occurs frequently in refactored codes. 
This phenomenon primarily stems from excessive method length, where code entanglement prevents clean extraction.
}
(2) \textbf{The LLM's performance decreases as the code complexity increases.} 
\tool attempts to reduce the cognitive complexity by splitting a function into multiple smaller functions. 
However, correctly and effectively decomposing a complex function is a non-trivial task, as it requires an in-depth understanding of the programming context, such as the function design, function logic and its interaction with other parts of the code. 
This is the reason why our approach performs suboptimal when dealing with \textit{Long Parameter List} and \textit{High Cognitive Complexity} smells \xt{through method decomposition}. 
How to better utilize LLMs to address code smells of complex code is interesting, but it is beyond the scope of our research.
(3) \textbf{Then LLM's hallucination problem.} 
As shown in Fig.~\ref{fig:fail_case}, when \tool extracts a code snippet into smaller functions, it mistakenly replaces the original code \texttt{atr[idx[0]] = element} with \texttt{atr[idx] = element}, which alters the functionality of the code after refactoring. 
However, only one participant in our user study found this functional inconsistency. 
This suggests that the hallucination problem poses threats to refactoring code with LLMs, developers must carefully review the refactored code to avoid potential bugs. 

\begin{figure}
    \centering
    \includegraphics[width=1.0\textwidth]{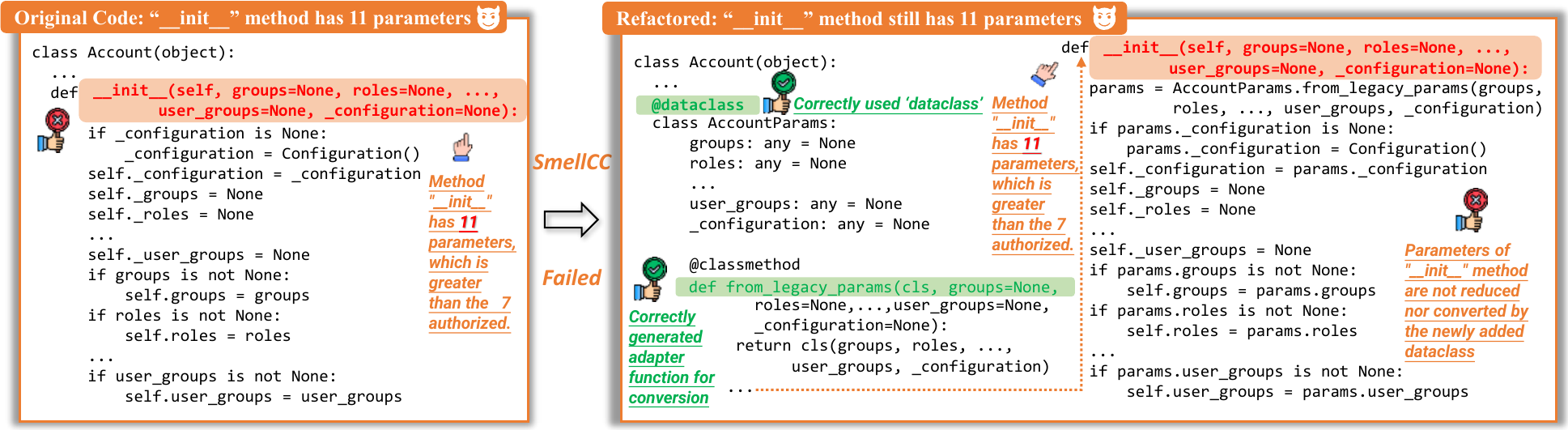} 
    \caption{A Failed Long Parameter List Case}
    \label{fig:fail-long-param-list}
\end{figure}

\xt{
As for the \textit{Long Parameter List} smell, 
to better illustrate the technical challenges involved, Fig.~\ref{fig:fail-long-param-list} presents a representative case, where the \texttt{\_\_init\_\_} method of the \texttt{Account} class initially accepts 11 parameters. 
While \tool correctly identifies the smell and applies the refactoring strategy — introducing a dataclass (i.e., \texttt{AccountParams}) to encapsulate parameters and generating an adapter method (i.e., \texttt{from\_legacy\_params}) for backward compatibility — the refactoring ultimately fails to reduce the number of parameters in the target \texttt{\_\_init\_\_} method.
The root cause lies in \tool's inability to fully propagate the parameter consolidation to the critical call sites and ensure all callers adopt the new parameter structure. 
In the refactored code, the \texttt{\_\_init\_\_} method retains all original parameters (e.g., \texttt{groups}, \texttt{roles}, \texttt{user\_groups}) despite the newly introduced \texttt{AccountParams} abstraction. 
Consequently, the refactored code merely shifts parameter grouping to the adapter method without simplifying the interface of \texttt{\_\_init\_\_}, leaving the core smell unresolved. 
This pattern accounts for a significant proportion of failed cases in our experiment. 
The tool’s reliance on localized transformations—without global analysis of method invocations—limits its ability to enforce holistic parameter reduction.
This pattern accounts for a significant proportion of failed cases in our experiment, as \tool's reliance on localized transformations - without global analysis of method invocations - limits its ability to enforce holistic parameter reduction.

\find{\textbf{Summary 2:} Our \tool performs localized syntactic transformations (e.g., method decomposition or parameter encapsulation) but lacks global program analysis to ensure holistic refactoring effectiveness, particularly in complex cases.}

\subsection{Implications for Research and Practice}
To advance both theoretical understanding and practical applications of LLM-assisted refactoring, we derive several actionable implications from our study's findings for: 
(1) advancing research methodologies in AI-assisted software maintenance, and (2) optimizing industrial adoption of LLM-based refactoring tools. 
These insights emerge from three key empirical relationships: (a) the propagation of code smells from training data to model outputs, (b) the differential efficacy of LLMs across smell types, and (c) the measurable impact on downstream tasks. 

\subsubsection{For Software Engineering Researchers.}
\textbf{Implication 1: Dataset quality standards should systematically incorporate code smell metrics.} 
Our findings in RQ1 and RQ3 reveal that more than 85\% of code smells in \texttt{CodeSearchNet}-Python propagate to LLM-generated outputs, while smell removal improves downstream tasks' performance (RQ4). 
This demonstrates that code smells constitute a significant but understudied dimension of dataset quality. 
We recommend prioritizing code smell metrics (e.g., smell density, type distribution) during dataset construction and evaluation.
\textbf{Implication 2: LLMs represent a new paradigm for automated refactoring when properly augmented with reasoning frameworks.} 
Our results demonstrate that LLMs equipped with Chain-of-Thought (CoT) prompting and few-shot learning achieve 96.8\% smell elimination while preserving functionality in 91.3\% of cases - performance surpassing traditional rule-based tools for most smell categories. 
This capability stems from LLMs' unique capacity to simultaneously analyze code structure, infer programmer intent, and generate syntactically valid transformations. 
However, the technology's current limitations in handling architectural smells like \textit{Long Parameter List} (60.9\% efficacy) reveal the need for hybrid approaches. 
Future systems should combine LLMs' pattern recognition strengths with static analyzers' precision for cross-file dependency resolution, creating a new class of AI-assisted refactoring tools that leverage the complementary strengths of both techniques.
\textbf{Implication 3: Code smells function as latent distractors that systematically impair LLMs' learning effectiveness.} 
Our RQ4 results demonstrate that models trained on smell-contaminated datasets exhibit poorer performance in both code completion and code search tasks compared to those trained on smell-free data.
This consistent performance degradation across multiple downstream tasks suggests that code smells introduce attention misallocation during model training, leading to suboptimal feature learning. 
These findings establish that conventional "correctness-only" data cleaning is insufficient - comprehensive dataset purification must proactively eliminate code smells to ensure both functional validity and learning-process purity.

\subsubsection{For Industry Practitioners.}
\textbf{Implication 1: Real-time smell prevention should integrate with development workflows.} 
The predominance of preventable \textit{Naming Convention} smells (74.39\% of refactored smells) suggests current development practices insufficiently address code quality at creation time. 
Our user study reveals developers typically identify these smells only during later reviews, resulting in higher remediation costs compared to immediate fixes. 
\textbf{Implication 2: Human oversight remains critical for complex refactorings.}  
Our experiments reveal clear limitations in fully automating architectural changes: while \tool handles simple refactorings like renaming with 91.4\% accuracy, it makes mistakes in 27.6\% of complex cases like simplifying \textit{Long Parameter List}s - especially when changes span multiple files. 
The solution isn't abandoning automation, but smart collaboration: let the tool flag issues and suggest fixes, then have senior engineers review system-wide impacts before finalizing changes.
\textbf{Implication 3: Risk-stratified refactoring workflows optimize human-AI collaboration efficiency.} 
Our experimental results in RQ2 demonstrate that LLM-assisted refactoring efficacy varies significantly by smell complexity, enabling a three-tiered approach: (1) High-confidence fixes (e.g., six perfect-refactoring smell cases) can be fully automated; (2) Medium-risk changes (e.g., \textit{High Cognitive Complexity} with 88.4\% accuracy) can benefit from LLM-generated drafts with developer validation; (3) High-risk architectural smells (e.g., \textit{Long Parameter List}s with 72.4\% accuracy) can use LLM suggestions as inspiration for manual refactoring. This workflow balances automation benefits (96.8\% overall efficacy) with quality control (91.3\% functional consistency).
}
\section{THREATS TO VALIDITY}
\noindent \textbf{Internal Validity.} We acknowledge that there may be potential bias in selected representative examples for the few-shot learning of \tool.  We invite two developers, with over 6 and 9 years of Python programming experience, to manually fix the 10 different code smells respectively and discuss with the first author to select several representative datasets. 
Another internal validity is the replication and fine-tuning of LLMs. Our experiment employed based on official sources from reputable AI communities like HuggingFace~\cite{huggingface}, and fine-tuned the DeepSeek-Coder following the implementation provided in their GitHub project~\cite{DeepSeek-Coder_Project} and adopted the recommended hyperparameter settings.

\noindent \textbf{External Validity.} There may be generalizability issues and selection bias regarding the applicability training dataset and LLMs used in the study.
We mitigated such issues by utilizing the representative dataset and LLMs. The selected \texttt{CodeSearchNet} is widely used by several LLMs and the selected LLMs are popular in solving software engineering tasks. 
\xt{While our experiments use the widely-adopted \texttt{CodeSearchNet}-Python benchmark, its known limitations — including (1) skewed language distribution, (2) unfiltered code noise, and (3) temporal misalignment of coding patterns — may pose threats to the generalizability of our findings.
In the future, we plan to extend our study to more benchmark datasets and LLMs.
Differences in programming languages may also be an external threat. 
While our \tool demonstrates efficacy in Python, its extension to other languages may require adaptations due to paradigm-specific syntax and idiomatic constraints. 
The current Chain-of-Thought decomposition and few-shot examples are optimized for Python code refactoring, and their direct application to languages with distinct paradigms could necessitate revised refactoring templates - a direction we prioritize in future work.
Although \tool demonstrates LLM-based refactoring viability, its advantages over rule-based tools lack systematic benchmarking. Future work will include systematic benchmarks against other refactoring baselines to quantify LLM advantages where traditional methods fail.
}
The final external threat is that our \tool is limited to handle ten types of code smell which is most frequent based on our preliminary study. In the future, we will extend \tool to more code smells.
\section{RELATED WORK}
\subsection{Dataset Quality for Code Large Language Model} 
In recent years, researchers have reached a consensus on the principle of \textit{garbage in, garbage out}~\cite{Sanders2017GARBAGEIG}, emphasizing that the quality of input data significantly affects the performance of any data-driven system, including code-based LLMs~\cite{jain2024llmassisted,DBLP:journals/corr/abs-2401-14196, 10597991}. To build high-quality datasets, considerable efforts have been devoted to enhancing various characteristics, such as readability~\cite{soldaini2024dolma, jain2024llmassisted, mai2024human, xiang2025automating, hu2021automating}, correctness~\cite{DBLP:journals/corr/abs-2401-14196, dai2025less, liu2024aw4c}, consistency~\cite{shi2022we, mai2025towards, gao2021automating}, and diversity~\cite{wei2023magicoder, luowizardcoder, dai2024mpcoder}.

For instance, Jain et al.~\cite{jain2024llmassisted} improved dataset readability by automatically modularizing long methods and providing detailed explanations for each smaller code segment, which further enhanced code generation performance. In efforts to ensure correctness, DeepSeek-Coder~\cite{DBLP:journals/corr/abs-2401-14196} employed a compiler and a quality model, supplemented by heuristic rules, to filter out low-quality data, including code with syntax errors. DeepSeek-Coder achieved an impressive improvement in code generation after training on the filtered dataset.
Furthermore, Shi et al.~\cite{shi2022we} explored the impact of semantic consistency between code and descriptions, demonstrating that correcting inconsistent code-description pairs leads to improved code summarization performance. 
Finally, WizardCoder~\cite{luowizardcoder} utilized Evol-Instruct technology to generate variable code questions and corresponding solutions to improve the diversity of the dataset. Fine-tuning on this expanded dataset endowed WizardCoder with remarkable code generation capabilities.

In this study, we focus on improving the quality of dataset in terms of code smell. We found that the code smells extensively exist in LLMs’ training datasets. We proposed an LLM-based tool to eliminate the code smell in training datasets. We found that the smell-clean dataset can facilitate LLM to produce high-quality code, as well as improve performance in downstream tasks, such as code generation and code search.

\subsection{Code Smells Refactoring}
A code smell is an indication of potential issues in the code's design or structure, these issues may not directly cause bugs but can negatively impact readability, maintainability, and scalability~\cite{lacerda2020code}. 
Refactoring is a practical approach to eliminate the code smell, reduce its negative impact, and further improve the maintainability of projects~\cite{opdyke1992refactoring, 1265817, fowler2018refactoring}. 
However, manual refactoring requires sufficient experience in the development and knowledge of the project, which is difficult and time-consuming~\cite{lacerda2020code, xing2006refactoring, wang2024insights, mai2025code2api}.
To this end, researchers have paid more attention to automatically refactoring the code smell issues~\cite{baqais2020automatic, naik2024deep, nyirongo2024survey}.

Nowadays, there are already numerous tools for automatically eliminating code smells, such as Rope~\cite{rope} for Python, JDeodorant~\cite{tsantalis2017clone} for Java, and CppRefactory~\cite{cpprefactory} for C++. 
Some tools, whether commercial or academic, also have been proposed to support refactoring in the form of IDE plugins~\cite{lacerda2020code, soares2010making, terra2018jmove}. 
However, current automated refactoring tools are designed based on carefully designed rules, which lack the capability to address higher-level design issues. They focus primarily on low-level code smells such as unifying format without considering context and program design. 
Recently, a considerable amount of research~\cite{naik2024deep, nyirongo2024survey, dong2024context, chi2023automated, liu2019deep} has been proposed to investigate and explore the application and adoption of deep learning techniques to support the process of software refactoring. 
For example, Ma et al.~\cite{ma2023pre} employed the pre-trained model CodeT5 to extract the semantic relationship between code 
snippets to detect feature envy code smell, providing the recommended refactoring solutions for the code smell. Tufano et al.~\cite{tufano2019learning} regarded code refactoring as the translation task, which trained a Neural Machine Translation (NMT) model to learn how to apply code changes implemented by developers automatically. Inspired by the strong capability of LLM to assist developers in daily development, some researchers proposed LLM-based refactoring technologies. For example, EM-Assist~\cite{wadhwa2024core} focuses on resolving \textit{Long Parameter List} issue, which validates, enhances, and ranks the refactoring candidates before presenting the results to the developer.

In this study, we proposed an LLM-based smell cleaning tool, named \tool, which utilizes the Chain-of-Thought strategy and few-shot learning to remove code smells in the given method. Our experiment indicates that \tool can effectively clean the top-10 frequent code smells detected by SonarQube and construct a high-quality smell-clean dataset.

\section{CONCLUSION AND FUTURE WORK}
In this study, we construct a systematic research to assess and improve dataset quality in terms of code smells. 
Firstly, our preliminary study shows that code smells extensively existed in benchmark datasets and they can indeed affect the quality of the LLM generated code. 
Afterward, we propose an LLM-based code smell cleaning tool, named \tool, which automatically refactors and removes code smells for a given method. 
The automatic evaluation and manual analysis verify the effectiveness of our tool for cleaning code smells. 
What's more, the correctness verification indicates that our \tool can remove most of the code smells while maintaining the same functionality.
We then apply our tool on \texttt{CodeSearchNet}-Python dataset to eliminate potential code smells and curate a smell-cleaned dataset. 
By fine-tuning with the smell-clean dataset, the LLM can reduce code smells in its generated code and improve the output quality.
\xt{The smell-clean dataset can also enhance the model’s performance in code completion and code search tasks.
Lastly, we derive several actionable implications for software engineering researchers and industry practitioners from our findings.
} 
In the future, we plan to enhance \tool by expanding its capabilities to identify and eliminate a broader range of code smells and more benchmark datasets.

\section{DATA AVAILABILITY}
Our replication package is available at~\cite{replication_package}

\bibliographystyle{ACM-Reference-Format}
\bibliography{software}

\end{document}